\documentclass[]{aa} 
\titlerunning{Discovery of HEC14}
\authorrunning{J. D. Sakowska et al}
\usepackage[utf8]{inputenc}

\usepackage{natbib}
\bibpunct{(}{)}{;}{a}{}{,}
\usepackage{gensymb}
\usepackage{caption}

\usepackage{pdflscape}
\usepackage{longtable}

\usepackage{graphicx}
\usepackage{txfonts}
\usepackage[colorlinks=true, linkcolor=blue, citecolor=blue, urlcolor=blue]{hyperref}

\begin{document}

    \title{Discovery of HEC14: an ultra-faint stellar system in M31’s outer halo at the globular cluster--dwarf galaxy boundary}

   \author{Joanna D. Sakowska
          \inst{1}\fnmsep\thanks{Corresponding author: \url{jsakowska@iaa.csic.es}},
          David~Mart\'inez-Delgado\inst{2,3}\fnmsep\thanks{ARAID fellow}, 
          Massimiliano Gatto\inst{4}, 
          Matteo Monelli\inst{5,6,7}, Giuseppe Donatiello\inst{8}, 
          Andrés del Pino\inst{1},
          Michelle L.~M. Collins\inst{9},
          Amandine Doliva-Dolinsky\inst{9},
          Ethan D. Taylor\inst{10}
          }
\institute{
$^{1}$ Instituto de Astrofísica de Andalucía (CSIC), Glorieta de la Astronom\'ia s/n, 18008 Granada, Spain\\
$^{2}$ Centro de Estudios de F\'isica del Cosmos de Arag\'on (CEFCA), Unidad Asociada al CSIC, Plaza San Juan 1, 44001 Teruel, Spain\\
$^{3}$ ARAID Foundation, Avda. de Ranillas, 1D, E-50018 Zaragoza, Spain\\
$^{4}$ Osservatorio Astronomico di Capodimonte, Napoli, Salita Moiariello, 16 - 80131 Napoli, Italy\\
$^{5}$ INAF-Osservatorio Astronomico d'Abruzzo, via Mentore Maggini s.n.c., 64100 Teramo, Italy \\
$^{6}$ IAC- Instituto de Astrof\'isica de Canarias, Calle V\'ia Lactea s/n, E-38205 La Laguna, Tenerife, Spain\\
$^{7}$  Departmento de Astrof\'isica, Universidad de La Laguna, E-38206 La Laguna, Tenerife, Spain \\
$^{8}$UAI -- Unione Astrofili Italiani /P.I. Sezione Nazionale di Ricerca Profondo Cielo, 72024 Oria, Italy \\
$^{9}$ Department of Physics, University of Surrey, Guildford GU2 7XH, UK \\
$^{10}$ Institute for Computational Cosmology, Department of Physics, Durham University, South Road, Durham DH1 3LE, UK\\
}

    \date{Received 18th November 2026; accepted XX XX 2026}

    \abstract
   {}
   {We present the discovery of HEC14, an ultra-faint stellar system near M31, visually identified in the Pan-Andromeda Archaeological Survey and characterised with deeper OSIRIS+@GTC imaging.}
   {HEC14's colour-magnitude diagram shows a clear red giant branch and a well-populated blue horizontal branch, indicating old (>10 Gyr), metal poor stellar populations. We find 912$^{+88}_{-80}$ kpc to be the best-fitting heliocentric distance to the satellite, consistent with M31's outer halo.}
   {With $M_{V} \sim -5.00^{+0.51}_{-0.44}$ and a half-light radius of $r_{\rm h} \sim 31^{+13}_{-9}$pc HEC14 lies in the size-luminosity gap between globular clusters and dwarf galaxies. The satellite therefore sits in the extended clusters (EC) regime, making it one of the faintest ECs discovered yet in M31.}
   {HEC14 is a new member of M31's halo EC population, whose origins remain unclear two decades after their discovery. We serendipitously find that HEC14's size and luminosity match theoretical predictions for dwarf galaxies with globular cluster-like sizes. Star formation history and/or spectroscopic studies are required to determine whether HEC14 (and possibly all ECs) is genuinely an extended globular cluster or a compact dwarf galaxy.}


   \maketitle
    
\nolinenumbers

\section{Introduction}

\begin{figure*}
\begin{center}

\includegraphics[width=1\textwidth]{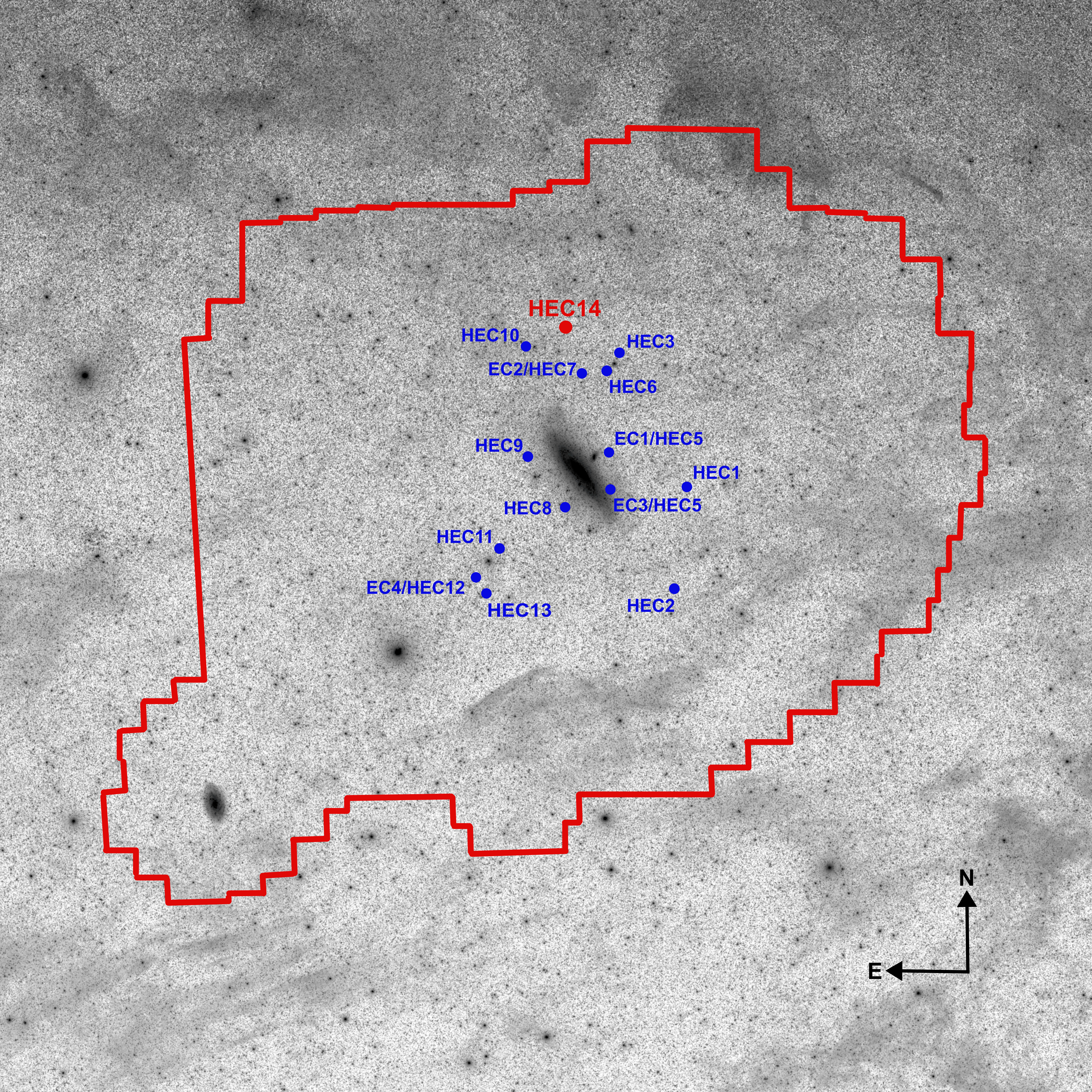}
\caption{HEC14's location (red spot) within the PAndAS survey footprint (\citealt{McConnachie2018}). At M31's distance (776 kpc) HEC14 is located approximately $\sim$61 kpc in projected distance from M31's centre. In blue we overplot the locations of all known halo ECs (\citealt{Huxor2011}). Image credits: And450 ultra-deep astrophotography survey (\citealt{Donatiello2025}).}
\label{fig:footprint}
\end{center}
\end{figure*}

\begin{figure*}
\begin{center}
\includegraphics[width=1\textwidth]
{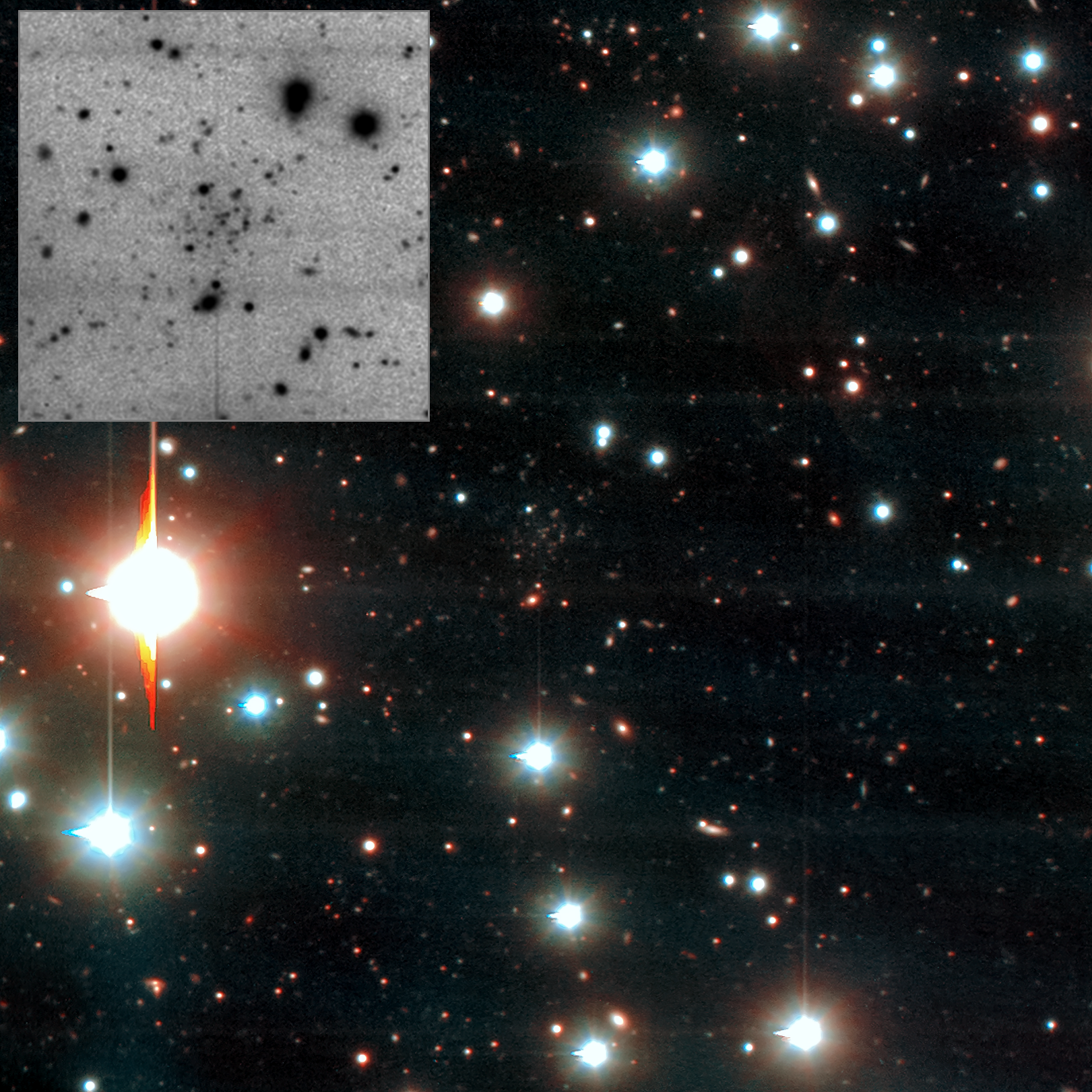}
\caption{Stacked GTC image with a negative colour zoom-in on HEC14 in the centre. North is up and east is left.}
\label{fig:GTC}
\end{center}
\end{figure*}

Historically, dwarf galaxies and globular clusters (GCs) were considered two distinct classes with little debate on their definitions. Dwarf spheroidal galaxies (dSphs) were regarded as extended systems (half-light radius $r_h > 100$ pc) formed in the early Universe, with prolonged star formation histories (SFHs) and high dynamical-to-stellar mass ratios owing to the presence of a dark matter (DM) halo (e.g., \citealt{Aaronson1983, Grebel1997}). GCs, on the other hand, were viewed as compact ($r_h \sim 1 - 10$ pc) low-mass systems composed of a chemically homogeneous stellar population, with light element abundance variations unique to GCs only (see \citealt{Kraft1994} and references therein), showing no dynamical evidence for DM (e.g., \citealt{Illingworth1976, Peterson1986}). Although the two classes span similar luminosities, measuring their apparent sizes was typically sufficient to cleanly separate the two in the luminosity-size plane.

The first challenge to these definitions came from the discovery of ultra-faint dwarf galaxies (UFDs) in the Sloan Digital Sky Survey (SDSS; \citealt{SDSS}). These diffuse, unusually faint systems quickly filled the size gap between GCs and dSphs (e.g., \citealt{Willman2005a, Willman2005b, Belokurov2006, Zucker2006}) and it was not immediately clear that UFDs were galaxies until the presence of DM was spectroscopically confirmed (e.g. \citealt{SimonGeha2007}). UFDs are now known to be the least luminous, least massive ($M_{V}>$ --7.7, $M_{*} \leq 10^{5}M_{\odot}$; see \citealt{Simon2019} for a review) and the most DM-dominated galaxies to exist (\citealt{Bullock2017}). UFDs constitute the relics of the earliest galaxies having formed most of their stars before the epoch of reionisation (e.g.\citealt{Gnedin2006}) as reflected in their short SFHs and very low metallicities ([Fe/H] $\lesssim-2.2$; e.g., \citealt{Kirby2011, Weisz2026}). As the depth and sensitivity of wide-field sky surveys improved (e.g., PanSTARRS: \citealt{Chambers2016}; the Dark Energy Survey: \citealt{DES}; the DESI Legacy Survey: \citealt{Dey2019}; DELVE: \citealt{Drlica-Wagner2021,Drlica-Wagner2022}; UNIONS: \citealt{unions}) new discoveries of Milky Way (MW) UFDs and hyper-faint GCs rapidly populated the size-luminosity plane down to extreme luminosities of $M_{V} \sim$ 0 (\citealt{Pace2025}). Fainter than $M_{V} \sim -4$, however, UFDs are comparable in size to the most extended MW GCs ($r_h \lesssim 30$ pc) and spectroscopy is required to segregate these systems, sometimes with fewer than 10 stars to test for a metallicity spread or DM (e.g., \citealt{Cerny2026}). Each new UFD or ancient GC is therefore a valuable test of the role of DM in small-scale structure formation. 

We currently know of stellar systems down to $M_{V}\sim-4$ for our nearest massive neighbour, Andromeda. Wide-field surveys with 4-m class telescopes struggle to find satellites fainter than $M_{V}\sim-6$ given that, at the distance of M31 (776 kpc, \citealt{Savino2022}), these telescopes are typically able to resolve red giant branch (RGB) stars only. Such faint satellites usually have very few RGB stars, necessitating deeper follow up imaging for their characterisation with either 10-m class telescopes (which reach the horizontal branch, HB) or with very deep space photometry (down to the oldest main-sequence turnoff). The emergence of so-called "extended clusters" in M31 (ECs: \citealt{Huxor2005, Mackey2006, Huxor2011, Tanvir2012, Mackey2013b, Huxor2014}), populating the gap between GCs and UFDs ($15 \leq r_h \leq 35$ pc; $-8 \leq M_{V} \leq -5$), provoked special interest. First, while the faintest ECs overlap with the MW's Palomar-type GCs, no MW counterparts are known for the four brightest M31 ECs (EC1, EC2, EC3, EC4\footnote{We note that ECs1-4 were renamed to HEC5, HEC7, HEC4, HEC12 (Halo Extended Clusters) by \cite{Huxor2011} to reflect their location in the M31 halo.}). Second, it remains to be convincingly demonstrated that all 13 ECs are indeed GCs- only the brightest ECs have been shown to be predominantly old and metal-poor, with no metallicity spread detected (down to the HB; \citealt{Mackey2006}). Similarly, no kinematic study has yet ruled out the presence of DM (\citealt{Collins2009, Veljanoski2014}). 

In an effort to discover new Andromeda satellites we ran a search campaign using archival Pan-Andromeda Archaeological Survey data (PAndAS; e.g. \citealt{McConnachie2009, McConnachie2018}). PAndAS provides systematic, continuous coverage across the M31/M33 system ($\sim$150/$\sim$50 kpc, respectively) using the MegaCam on the Canada-France-Hawaii Telescope (CFHT), resolving RGB stars across M31/M33 and their satellites (e.g. figure 2 of \citealt{Martin2016b}). PAndAS discovered 60+ new halo GCs (e.g., \citealt{Huxor2014}) and 20+ dwarf satellites (e.g. \citealt{martin2006, ibata2007, McConnachie2008, Martin2009, Richardson2011, Martin2016b}), including some of the faintest UFDs below $M_V \sim -6$ (confirmed with deeper follow-up imaging: \citealt{And35, Smith2025, Sakowska2026}). Our approach has contributed Pisces VII (Psc VII; \citealt{MD2022, Collins2024}), Pegasus V (Peg V; \citealt{Collins2022}) and Andromeda XXXVI (\citealt{Sakowska2026}) to the M31 satellite system.

In this paper we present the discovery of Halo Extended Cluster 14 (HEC14), a new EC located in the remote outer M31 halo. Sect. \ref{sec:meth} details our deep, follow-up OSIRIS+@GTC imaging. In Sect. \ref{sec:res} we characterise HEC14 and, in Sect. \ref{sec:disc}, we discuss its nature.

\begin{figure*}
\begin{center}
\includegraphics[width=1\textwidth]
{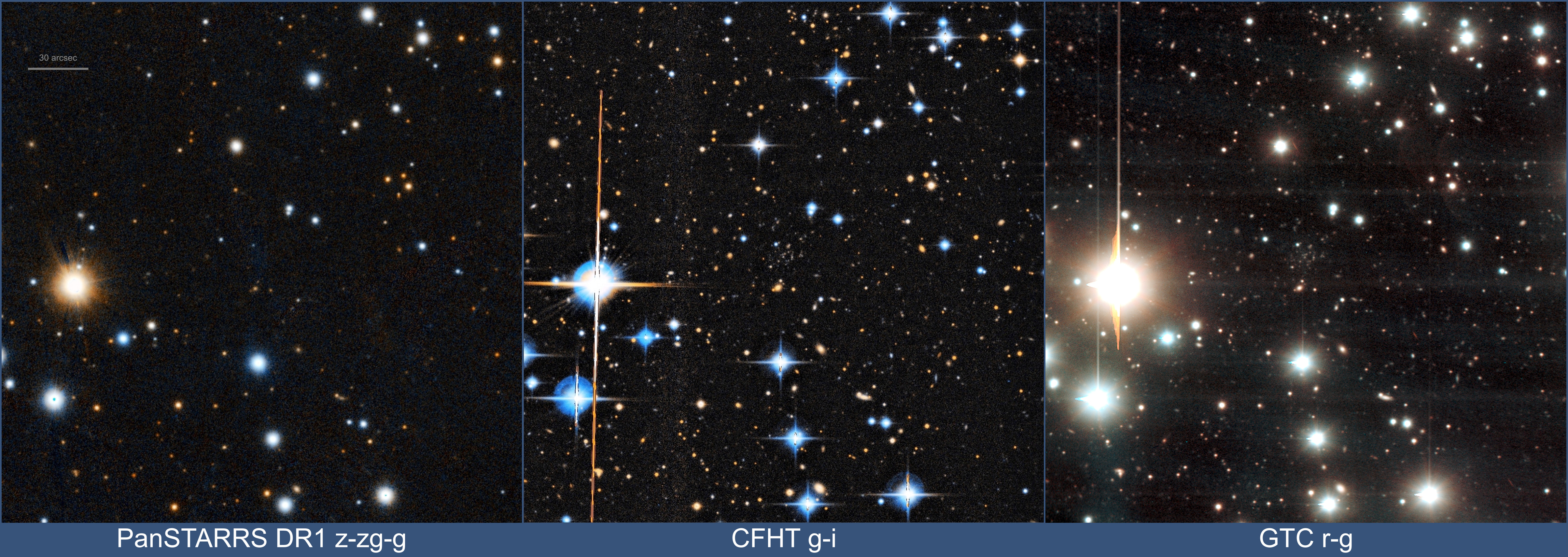}
\caption{Image of HEC14 across three different datasets. From left to right: PanSTARRS DR1, CFHT (PAndAS), and GTC (this work). HEC14 becomes visible only in CFHT and GTC data thanks to the increasing image depth. }
\label{fig:PAndAS}
\end{center}
\end{figure*}

\section{Data}
\label{sec:meth}

HEC14 was discovered by the amateur astronomer Giuseppe Donatiello, who identified 11 semi-resolved overdensities during a systematic visual inspection of public PAndAS survey stacked images across its full footprint (Fig.~\ref{fig:footprint}). The same visual approach has previously uncovered more than ten UFDs in the DESI Legacy Survey and ultra-deep imaging (M31 satellites: \citealt{MD2022, Collins2022}; Do I: \citealt{MD2018}; NGC 253 satellites: \citealt{MD2021, MD2025}). To confirm the two most promising candidates, we obtained Director's Discretionary Time (GTC2025-292.227, PI: Mart\'inez-Delgado) with the 10.4-m Gran Telescopio Canarias (GTC; Roque de los Muchachos Observatory, La Palma, Spain), which enables deeper imaging than with the 3.6-m CFHT. The first candidate was confirmed as the ultra-faint dwarf galaxy And XXXVI (\citealt{Sakowska2026}). Observations of the second candidate (HEC14) were taken at low elevation in late January 2026 and considerably affected by moonlight contamination. We re-observed HEC14 on 8 September 2026 as part of the 2026B program (GTC70-26B; PI. Martínez-Delgado) and use this dataset only.

To characterise HEC14 we used the Optical System for Imaging and low-Intermediate-Resolution Integrated Spectroscopy instrument (OSIRIS+)\footnote{\url{http://www.gtc.iac.es/instruments/osiris+/osiris+.php}} installed on the GTC. OSIRIS+ has an unvignetted field of view of 7.8$'\times$7.8$'$ and a pixel scale of 0.254$''$~pixel$^{-1}$. For optimum image sampling and minimisation of chip defects we employed a ten-point dithering pattern (10$\arcsec$ offsets). The total exposure times were 44 $\times$ 180\,s = 7920\,s in $g'$ and 40 $\times$ 100\,s = 4000\,s in $r'$, with a seeing of 0.7$''$ in both bands. We reduced the data with the SAUSERO pipeline\footnote{\url{https://zenodo.org/records/15855515}}, the recommended tool within the standard GTC pipeline\footnote{\url{https://www.gtc.iac.es/observing/datareduction.php}} for OSIRIS+ broadband imaging. The reduction comprised overscan correction, bias subtraction, flat-field division, and astrometric calibration against Gaia DR3 stars, followed by stacking to produce the final $g'$ and $r'$ images. Figures~\ref{fig:GTC} and~\ref{fig:PAndAS} show the final stacked HEC14 image- its stars are now better resolved than in the PAndAS (CFHT) or Pan-STARRS survey data.

Photometry was performed using the \verb|DAOPHOT| suite \citep{stetson87} based on the procedures described in \citealt{Monelli2010}: (i) aperture photometry on all sources (3$\sigma$ threshold), (ii) point spread function (PSF) estimation and (iii) PSF photometry using the \verb|ALLSTAR| code on individual images, (iv) \verb|ALLFRAME| on all images using a clean input catalogue derived on the stacked image. The PSF estimation (step ii) is automatic, using routines which detect bright sources with low photometric error in every frame, and accounts for spatial variations across the image. Background galaxies and unresolved sources were manually removed from the \verb|ALLFRAME| input catalogue (step iv). To calibrate the catalogue to PanSTARRS we assumed $g\_{cal} = g\_{instr} + 8.856 + 0.043*(g-r)\_{instr}$ and $r\_{cal} = r\_{instr} + 8.884 + 0.088*(g-r)\_{instr}$, and rms $\sim$0.04 mag. To correct for extinction we used the \citealt{Schlegel1998,
Schlafly2011} reddening map. For our star galaxy separation we selected sources measured to have \verb|SHARP|$< |0.2|$.

\section{Characterising HEC14}
\label{sec:res}          

\begin{figure*}
\begin{center}
\includegraphics[width=0.5\textwidth]{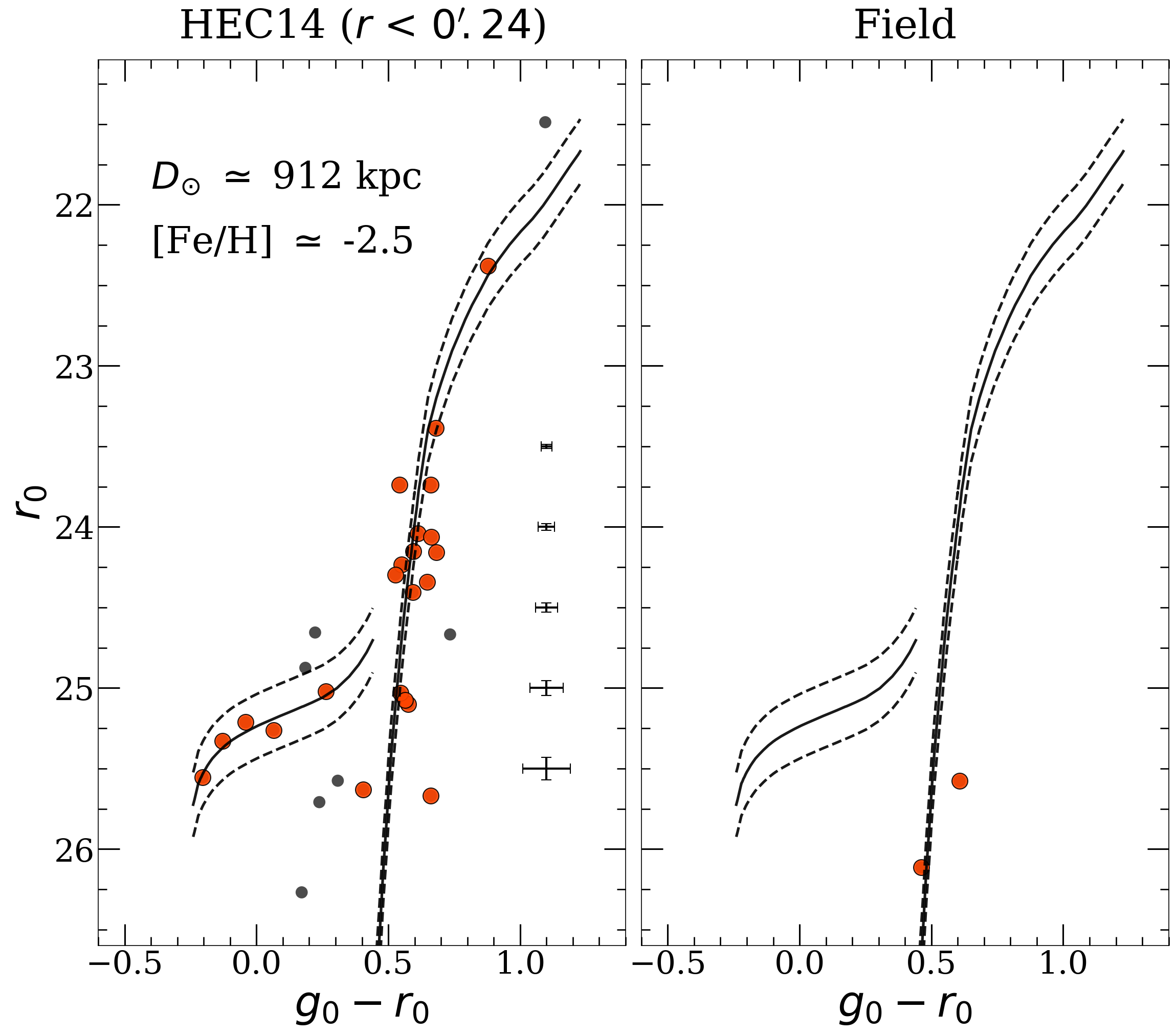}
\includegraphics[width=0.43\textwidth]{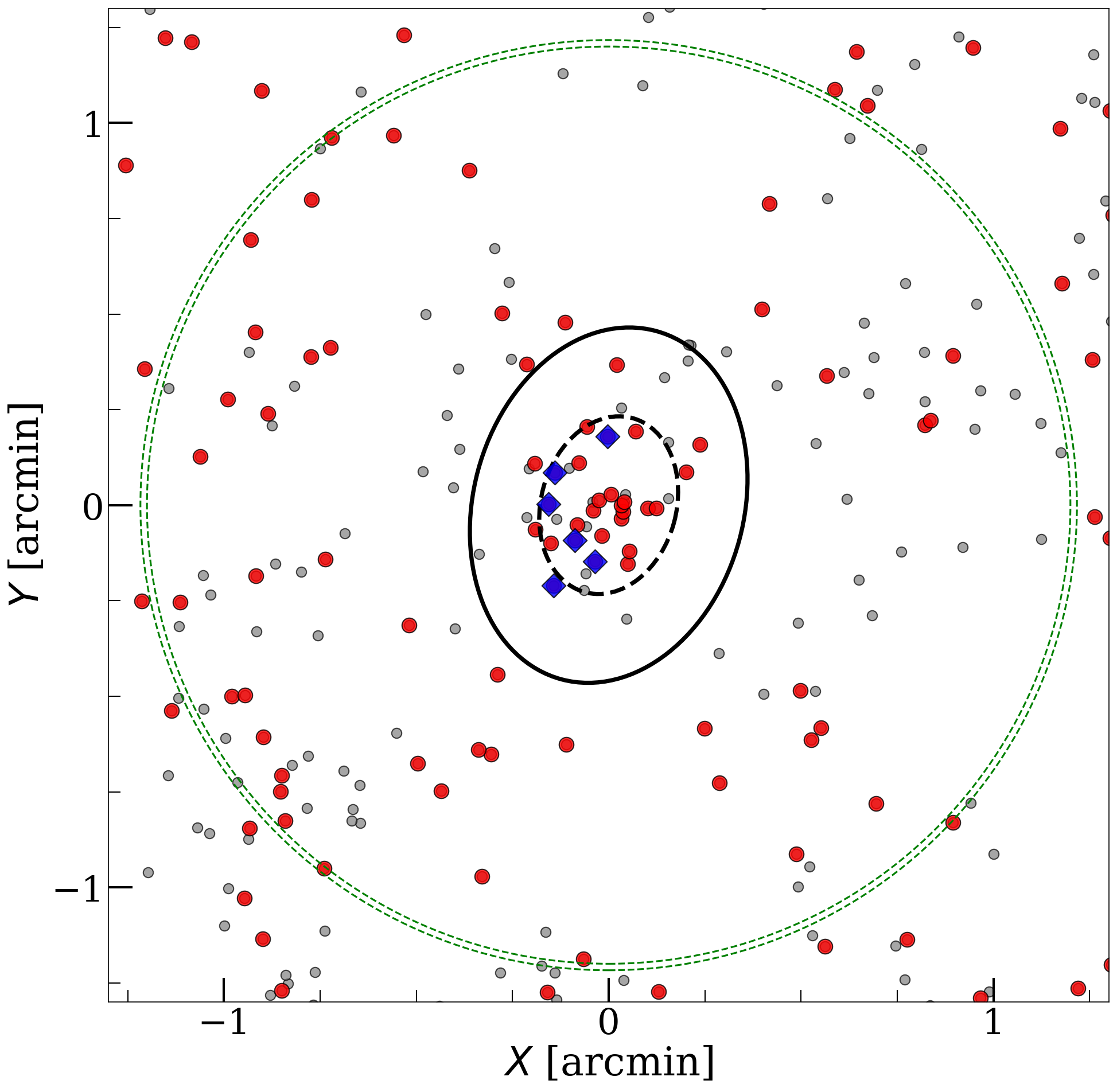}
3\caption{ HEC14 (left) colour-magnitude diagram (CMD) within 2$r_{h}$ (0.24 arcmin) of the overdensity's centre. Using a black solid line we overlay a 12.5 Gyr, [Fe/H] = $-$ 2.5 PARSEC isochrone at 912 kpc, demonstrating that the stellar content is compatible with being old and metal-poor. The red points correspond to the sources that have been isochrone-filtered and used for the structural analysis, while the grey points show the remaining sources classified as stars with our photometric selection cut. The dashed and dash-dotted line mark the best fit isochrone shifted in distance modulus by -0.2 and +0.2 magnitudes, respectively. In the middle panel we show the CMD representing Field contamination. Red giant branch and horizontal branch stars are present in the HEC14 CMD and are not observed in the Field. Right: 2D position map of all sources (grey) and isochrone filtered stars (red)- HEC14 is clearly visible as an overdensity. Using green annuli we mark the area chosen to represent Field contamination, and the area is equivalent to an ellipse of semi-major axis 2$r_{h}$. We plot these stars in the Field CMD (note that some stars fall outside of the CMD axis limits). We also overplot two black ellipses which have a semi-major axis equal to  2$r_{h}$ and 4$r_{h}$, respectively, reflecting the ellipticity and position angle derived in the analysis (see text for details). Using blue diamonds we highlight horizontal branch stars, demonstrating that all but one lie within 2$r_{h}$ and are therefore probable member stars. }
\label{fig:CMD}
\end{center}
\end{figure*} 

In Figure \ref{fig:CMD} we show HEC14's CMD on the left, choosing stars within $2r_{\rm h}$ (0.24 arcmin; determined in the next step). The CMD shows all sources after our star-galaxy separation cut. HEC14's CMD delineates a RGB and candidate HB stars, both of which are not seen in the control CMD (middle panel). The control CMD is made by placing an annulus centred on the overdensity (of radius 1.5 arcmin, see right panel) and using a second annulus to close the control region. The area of the control region is equivalent to an ellipse with a semi-major axis equivalent to $2r_{\rm h}$. HEC14's CMD shows four blue HB candidate stars within $2r_{\rm h}$ of its centre, three of which are located in a CMD region where no MW contamination is expected. The reddest HB star ($g_0 - r_0 = 0.2 - 0.3$) risks being a foreground star, given MW halo contamination can appear as an almost vertical CMD feature at $g_0 - r_0 = 0.2 - 0.4$. A blue HB morphology is correlated with an old (> 10 Gyr) relatively metal-poor stellar population (e.g. \citealt{McGill2025}, their fig. 4); with this information we visually overlapped a selection of PARSEC library (\citealt{Bressan2012}) isochrones, demonstrating that old (> 10 Gyr) and metal-poor ([Fe/H] < -- 2.0) isochrones fit the position of stars in the CMD well. As an example, in Figure \ref{fig:CMD} we overlap a 12.5 Gyr isochrone of metallicity [Fe/H] = -- 2.5, placed at a distance modulus of $m - M_{0}$ = 24.8 (912 kpc), finding a good overall fit to the RGB and HB. Dashed lines delineate the shifted isochrone corresponding to a conservative distance modulus uncertainty of $\pm$0.2 mag (+88 and -80 kpc, respectively), showing that it fully encompasses the HB stars. The isochrone and distance assumption will benchmark our subsequent analysis.

We applied a Markov chain Monte Carlo (MCMC) analysis to characterise the structural parameters of HEC14. This analysis is described in \cite{Martin2008, Martin2016b}, and has since been adapted by the team to successfully characterise UFDs (\citealt{MD2022, Collins2022, Collins2024,Gatto2024, Sakowska2026, Gatto2026}). In essence, the analysis constructs a likelihood function $\rho_{\rm model}(r)$ to describe the two-dimensional positions of stars in the image:

\begin{equation}
    \rho_{\rm model}(r) = \rho_{\rm system}(r) + \Sigma_{b} .
    \label{eq:total profile}
\end{equation}

where $\rho_{\rm system}(r)$ is the radial density profile of the stellar system) containing $N_{*}$ likely member stars, measuring a half-light radius $r_{\rm h}$ and ellipticity $\epsilon$. The radial density profile is a function of elliptical radius \textit{r}, which in itself depends on the system's central coordinates on the sky ($x_{0}$, $y_{0}$), the ($x$, $y$) coordinates of the tangent plane at field's centre, and the position angle of the major axis $\theta$. We approximate the background contamination $\Sigma_{b}$ as uniform, derived by subtracting stellar system member stars (measured by integrating $\rho_{\rm system}$) from the total number of identified candidate member stars.

To determine the model's structural parameters $x_{0}$, $y_{0}$, $r_{\rm h}$, $\epsilon$, $\theta$ and $N_{*}$ we ran an MCMC analysis using \verb|EMCEE| (\citealt{Foreman-Mackey2013}). We fed in stars only above our completeness magnitude cut $g_{0} < 25.5$, $r_{0} < 25$ and also those filtered by our best-fitting isochrone (red points; Fig \ref{fig:CMD}). For the MCMC priors we only set $\pm0.5$ arcmin for the central RA and Dec of the overdensity. We used 1000 walkers over a total of 10,000 iterations, with a burn-in stage of 1000. The maximum integrated autocorrelation time among the parameters is $\tau \approx 112$ steps, so the chains are $\approx 89\tau$ long and the burn-in corresponds to $\approx 9\tau$, ensuring convergence (\citealt{Foreman-Mackey2013}). In the Appendix Fig. \ref{fig:mcmc} we show the final output corner plot; the best-fitting structural parameters are measured from the posterior distribution's median, and their respective errors using the 16th and 84th quantile. 

\renewcommand{\arraystretch}{1.4} 
\begin{table}[h]
\centering
\caption{HEC14 properties determined from the MCMC analysis.}
\begin{tabular}{l c c}
\hline
Parameter & Value  \\
\hline
RA  & $00^{\mathrm h}46^{\mathrm m}31.0^{\mathrm s\,+0.12}_{-0.08}$ \\
Dec & $+45^\circ06'42''^{+1.8}_{-1.2}$ \\
$D_{\odot}$ (kpc) & $912^{+88}_{-80}$ \\
$r_h$ (arcmin) & $0.12 ^{+0.05}_{-0.03}$  \\
$r_{\rm h}$ (pc) & $31 ^{+13}_{-9}$ \\
$\epsilon$ & $< 0.44$\\
$\theta$ (deg) & $165^{+58}_{-48}$ &  \\
$N_\mathrm{*}$ & $13^{+4}_{-3}$  & \\
$M_{g}$ & $-4.50^{+0.51}_{-0.44}$   \\
$M_{r}$ & $-4.98^{+0.52}_{-0.43}$  \\
$M_{V}$ & $-5.00^{+0.51}_{-0.44}$  \\ 
$M_{*} (M_{\odot})$ & $9300^{+5200}_{-3800}$  \\
\hline
\end{tabular}\label{tab:mcmc_results}
\end{table}

\begin{figure*}
\begin{center}
\includegraphics[width=0.49\textwidth]{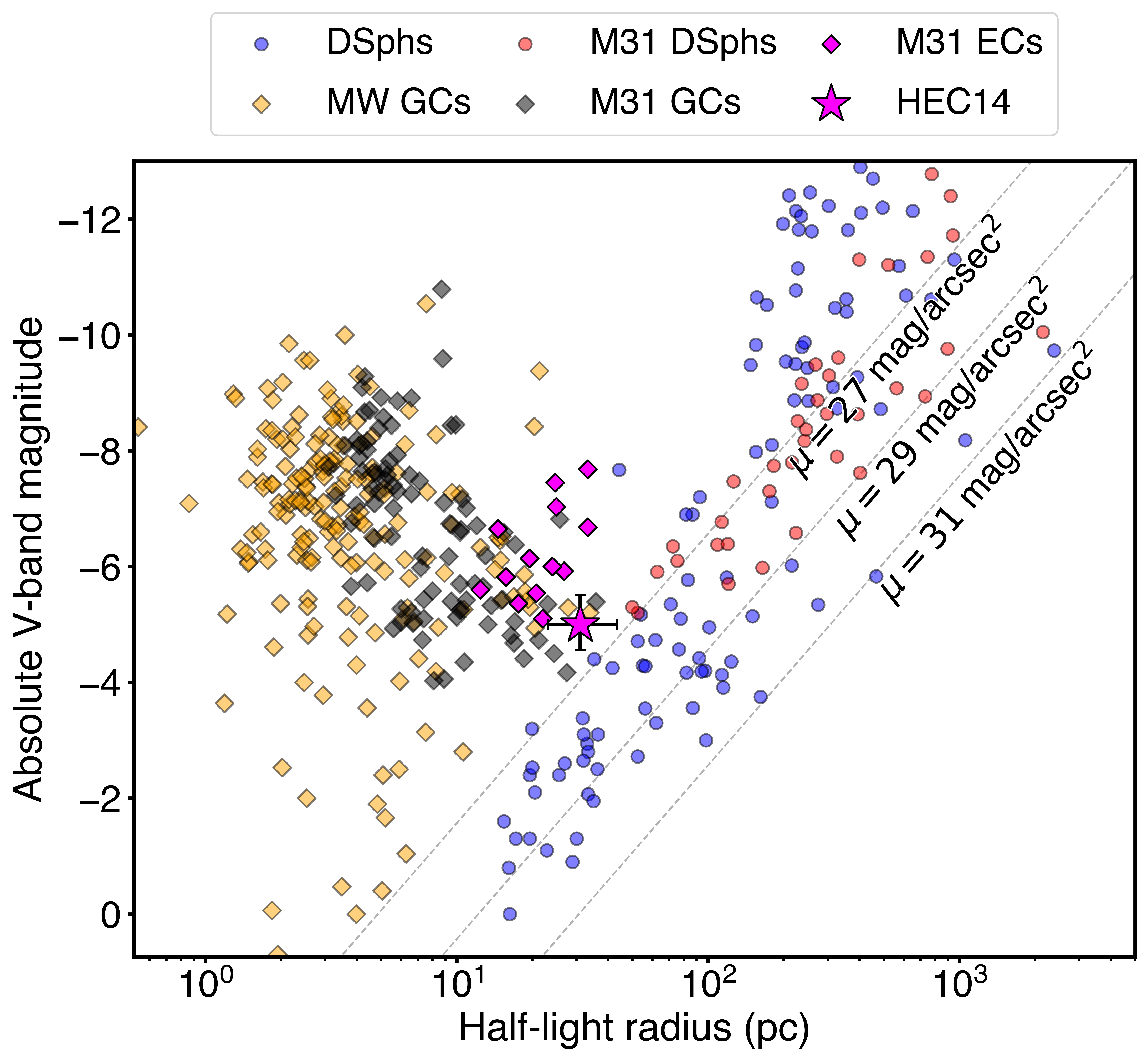}
\includegraphics[width=0.49\textwidth]{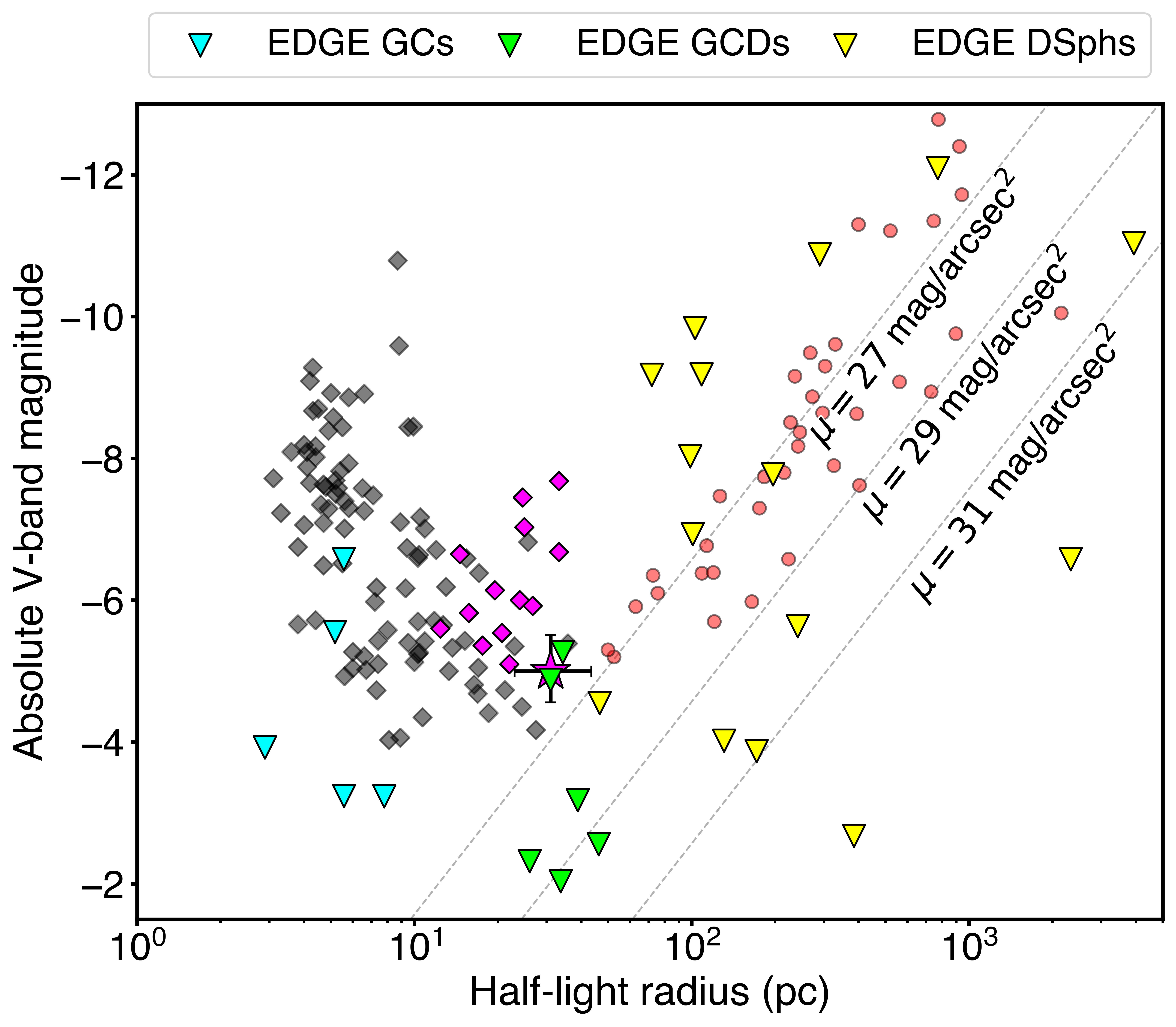}
\caption{Left: Absolute visual magnitude $M_{V}$ versus half-light radius $r_{h}$ (pc) for Milky Way (MW) and Local Volume dwarf spheroidal galaxies (dSphs), M31 dSphs, MW globular clusters (GCs), M31 GCs, M31 extended clusters (ECs) and HEC14. Right: Position of HEC14, M31 dSphs and M31 GCs against the \cite{Taylor2025} EDGE dwarf galaxy simulation predictions for GCs (blue), dwarf galaxies (yellow) and globular cluster-like dwarfs (GCD; green). Data taken from the Local Volume database (selecting only confirmed systems; \citealt{Pace2025}) and the PAndAS M31 halo GC catalogue (\citealt{Huxor2014}).}
\label{fig:Mv_rh}
\end{center}
\end{figure*}

In the right of Fig. \ref{fig:CMD} we show the 2D distribution of stars in the image and overlay two ellipses over HEC14, using a $2r_h$ and $4r_h$ semi-major axes and adjusted to reflect the ellipticity and position angle determined by \verb|EMCEE|. We also highlight locations of the HB stars, showing that four are located within $2r_h$ and one blue HB star is located just outside of $2r_h$. HEC14 is localised at RA = $00^{\mathrm h}46^{\mathrm m}31.0^{\mathrm s\,+0.12}_{-0.08}$ and Dec = $+45^\circ06'42''^{+1.8}_{-1.2}$, with a half-light radius of $r_h \sim 0.12 ^{+0.05}_{-0.03}$ arcmin, and the number of stars $N_\mathrm{*} \sim 13^{+4}_{-3}$. Although the MCMC analysis determined an ellipticity of $\epsilon \sim 0.26^{+0.21}_{-0.17}$, the posterior distribution of the ellipticity is not gaussian and is fully consistent with zero. We therefore quote $\epsilon < 0.44$ (at the 84th percentile) and note that the position angle of the major axis ($\theta \sim 165^{+58}_{-48}$ degrees) is likely unconstrained.

To measure the absolute V-band magnitude and observed stellar mass of HEC14, we continued the \cite{Martin2008, Martin2016b} approach and sampled a synthetic HEC14 CMD 10,000 times. The synthetic CMD is constructed by generating a $10^6M_{\odot}$ synthetic stellar population matching the PARSEC isochrone assumption (12.5 Gyrs, Fe/H = -- 2.5) and using a \citet{Kroupa2002} initial mass function (IMF). For each iteration, we drew $N_\mathrm{*}$ from a Poisson distribution centred on the observed number of stars, then randomly sampled stars from the synthetic CMD until the count of objects brighter than our completeness limit ($g_0 < 25.5$, $r_0 < 25$) reached $N_*$. We converted all of the $g, r$ stellar magnitudes (including those below the completeness cut) to a luminosity and stellar mass, summing to get the final average measurement for the system. We performed 10,000 iterations; in every iteration, we varied the distance assumption $(m - M_{0}) = 24.8$ (912 kpc) by drawing the distance modulus from a Gaussian distribution (centred on the distance) of width reflecting our distance uncertainty (0.2 magnitudes). We calculated $M_{g}\sim-4.50^{+0.51}_{-0.44}$ and $M_{r} \sim -4.98^{+0.52}_{-0.43}$, respectively, where the error is derived from the 16th and 84th percentiles of the MCMC analysis and accounts for the distance uncertainty ($24.8 \pm 0.2$ mag, equivalent to $912^{+88}_{-80}$ kpc). Applying the \cite{Jordi2006} transformation we obtain the absolute V-band magnitude as $M_{V} \sim -5.00 ^{+0.51}_{-0.44}$. Assuming a Kroupa initial mass function (\citealt{Kroupa2002}) the stellar mass of HEC14 corresponds to $M_{\star} \sim 9300^{+5200}_{-3800} M_{\odot}$. Assuming a distance of $D_{\odot} \simeq 912^{+88}_{-80}$, $r_h \sim 31^{+13}_{-9}$ pc. These results are summarised in Table \ref{tab:mcmc_results}.

\section{Discussion and conclusions}
\label{sec:disc}

In this work, we present the discovery of HEC14- a new ultra-faint stellar system located in the distant M31 halo. HEC14 was discovered as a semi-resolved overdensity in public PAndAS CFHT data and followed up for deeper imaging with the GTC. The resulting CMD revealed a clear RGB and a blue HB consistent with ancient, very-metal poor stellar populations; we demonstrated that an old (12.5 Gyr) very metal-poor ([Fe/H] = $-$ 2.5) isochrone is compatible with the blue HB (Fig. \ref{fig:CMD}) and fits it best when a heliocentric distance of 912$^{+88}_{-80}$ kpc is assumed. This distance assumption places HEC14 in the remote outer M31 halo (136$^{+88}_{-80}$ kpc behind M31) yet well within its virial radius ($r_{200}\sim 260$ kpc). Based on this, we characterised HEC14 to be extremely faint $M_{V}\sim- 5.00^{+0.51}_{-0.44}$, small $r_{\rm h} \sim 31 ^{+13}_{-9}$ pc and low in mass ($M_{\star} \sim 9300^{+5200}_{-3800} M_{\odot}$). In Fig. \ref{fig:Mv_rh} (left) we place HEC14 on the luminosity-size relation and compare it to M31/MW GCs and UFDs, finding that HEC14 sits in the gap occupied by M31 ECs. From a stellar populations perspective, the blue HB feature observed in HEC14 is also found in the brightest ECs (ECs1-4; \citealt{Mackey2006, Tanvir2012, McGill2025}) and their age-metallicity constraints are similar to those found for HEC14 ($\gtrsim$ 10 Gyrs old, $-2.2 \lesssim$ [Fe/H] $\lesssim - 1.8$; \citealt{Mackey2006}). We name our object HEC14 following the nomenclature of naming extended clusters located in M31's halo (\citealt{Huxor2011}). 

The formation mechanism behind ECs remains uncertain. Their blue HBs are compatible with both ancient GCs and UFDs (\citealt{Weisz2026}) and it is unclear whether these systems contain any DM (\citealt{Collins2009, Veljanoski2014}). If ECs are indeed GCs, then their ancient stellar populations, extended sizes, and location in the M31 halo could be explained with ex-situ origins. HEC14's structural parameters overlap with Sagittarius II and Palomar 14 in Fig \ref{fig:Mv_rh}, two metal-poor GCs in the MW outer halo which are consistent with having been accreted, and unlikely to contain significant DM (Palomar 14: \citealt{Sollima2011}, \citealt{Sollima2012}, \citealt{Caliskan2012}; Sagittarius II: \citealt{Laevens2015}, \citealt{Longeard2021}). Although many halo ECs have be found to be associated with M31 substructures (\citealt{Chapman2008, Forbes2010, Mackey2010}) not all of them are, or some associations are ambiguous (\citealt{Mackey2019}). Interestingly, \cite{McGill2025} showed a correlation between GC association with massive substructure in M31's halo and the presence of a red HB colour in their CMDs. This shows that halo GCs generally show younger and more metal-rich stellar populations, indicating a more recent formation time than would be compatible for ECs1-4 and HEC14 (blue HB stars) which would have formed earlier (either in-situ or ex-situ). From the PAndAS survey metal-poor RGB density map (\citealt{McConnachie2018}, their figure 12) HEC14 does not appear to be near major stellar substructure. However, a detailed study of this region using RGB density maps which vary in photometric metallicity (e.g. \citealt{Ibata2014}, their figure 9) is required to decisively probe whether minor stellar substructure could be present near HEC14.

Using the high-resolution EDGE dwarf galaxy simulations, \cite{Taylor2025} recently predicted a new class of DM-rich objects called globular cluster-like dwarfs (GCD). These objects form their stars in relatively small DM halos prior to re-ionisation ($M_{200} \sim 10^{6.3} - 10^{7.4} M_\odot$), contain very metal-poor stellar populations ([Fe/H] $< -2.5$), a metallicity spread larger than expected for GCs ($\sigma_{\rm [Fe/H]} \sim 0.2$) and a dynamical-to-stellar mass ratio between that expected for GCs and dwarf spheroidals ($\sim 10-50$). In Fig. \ref{fig:Mv_rh} (right) we overplot the size and luminosity predicted for GCDs finding that HEC14 could be a new candidate GCD within its error bars. Although challenging with current ground-based instrumentation, spectroscopically confirming HEC14's metallicity as below the GC metallicity floor at [Fe/H] $< -2.5$ (\citealt{Beasley2019, Kruijssen2019}) or even below [Fe/H] $< -2.2$ (expected for this luminosity; \citealt{Kirby2013}) would form a strong argument for a dwarf galaxy nature, and whether HEC14 could contain surviving metal-free stars (as predicted for GCDs). Constraining HEC14's metallicity, together with deriving its dynamical-to-stellar mass ratio, may prove an important probe into the existence of GCDs and possibly the nature of DM.

Deeper space observations of HEC14, with facilities such as the HST or Nancy Grace Roman Space Telescope\footnote{See the Roman Space Telescope Cycle 1 survey of M31/M33, which presents an excellent opportunity for not only characterising known ECs but also discovering new ones: \url{https://roman.ipac.caltech.edu/cycle1-approved-programs/2002}}, are essential to constrain HEC14's structural and luminosity parameters with higher precision. These facilities are also able to resolve stars down to the oldest main-sequence turn-off, enabling the derivation of quantitative SFHs with significantly reduced age-metallicity degeneracy (e.g., \citealt{Gallart2005}) key to confirming whether a metallicity spread exists in HEC14. Such a study, in addition to SFH studies of all known ECs, would be a valuable contribution to understanding whether ECs are truly extended GCs or, in fact, compact dwarf galaxies.

\begin{acknowledgements} Data are available upon request to the corresponding author. JS and AdP acknowledge financial support from RyC-MAX grant 20245MAX008 (CSIC) and the Severo Ochoa grant CEX2021-001131-S funded by MCIN/AEI/ 10.13039/501100011033. JS and DMD acknowledge financial support from project PID2022-138896NB-C53. DMD thanks financial support for a visiting researcher stay at the Astronomy and Astrophysics Department of the University of Valencia within the framework of the «Talent Attraction» programme implemented by the Office of the Vice-Principal for Research (INV25-01-15). MG acknowledges "Partecipazione LSST - Large Synoptic Survey Telescope" (Ob. Fu.: 1.05.03.06). MM acknowledges support from the Agencia Estatal de Investigaci\'on del Ministerio de Ciencia e Innovaci\'on (AEI-MCINN) under grants "At the forefront of Galactic Archaeology: evolution of the luminous and dark matter components of the MW and LG dwarf galaxies in the {\it Gaia} era" with references PID2020-118778GB-I00/10.13039/501100011033 and PID2023-150319NB-C21/10.13039/501100011033. AdP also acknowledges financial support from the Spanish Ministry of Science, Innovation and Universities project PID2024-155572NB-C22 (MICIU/AEI/10.13039/501100011033, FEDER, EU), the Ramón y Cajal fellowship RYC2022-038448-I (MICIU/AEI/10.13039/501100011033, co-funded by the European Social Fund Plus), and grant CNS2025-166660. MLMC and ADD acknowledge support from STFC grants ST/Y002857/1 and ST/Y002865/1. EDT acknowledges the Royal Society for funding their Royal Society Career Development Fellowship. Based on observations made with the GTC telescope, in the Spanish Observatorio del Roque de los Muchachos of the Instituto de Astrofísica de Canarias, under Director’s Discretionary Time GTC2025-292 and the 2026B semester with reference GTC70-26B.
\end{acknowledgements}

\bibliographystyle{aa} 
\bibliography{Bibliography}

\begin{thebibliography}{78}
\expandafter\ifx\csname natexlab\endcsname\relax\def\natexlab#1{#1}\fi

\bibitem[{{Aaronson}(1983)}]{Aaronson1983}
{Aaronson}, M. 1983, \apjl, 266, L11

\bibitem[{{Abazajian} {et~al.}(2009){Abazajian}, {Adelman-McCarthy}, {Ag{\"u}eros}, {Allam}, {Allende Prieto}, {An}, {Anderson}, {Anderson}, {Annis}, {Bahcall}, {Bailer-Jones}, {Barentine}, {Bassett}, {Becker}, {Beers}, {Bell}, {Belokurov}, {Berlind}, {Berman}, {Bernardi}, {Bickerton}, {Bizyaev}, {Blakeslee}, {Blanton}, {Bochanski}, {Boroski}, {Brewington}, {Brinchmann}, {Brinkmann}, {Brunner}, {Budav{\'a}ri}, {Carey}, {Carliles}, {Carr}, {Castander}, {Cinabro}, {Connolly}, {Csabai}, {Cunha}, {Czarapata}, {Davenport}, {de Haas}, {Dilday}, {Doi}, {Eisenstein}, {Evans}, {Evans}, {Fan}, {Friedman}, {Frieman}, {Fukugita}, {G{\"a}nsicke}, {Gates}, {Gillespie}, {Gilmore}, {Gonzalez}, {Gonzalez}, {Grebel}, {Gunn}, {Gy{\"o}ry}, {Hall}, {Harding}, {Harris}, {Harvanek}, {Hawley}, {Hayes}, {Heckman}, {Hendry}, {Hennessy}, {Hindsley}, {Hoblitt}, {Hogan}, {Hogg}, {Holtzman}, {Hyde}, {Ichikawa}, {Ichikawa}, {Im}, {Ivezi{\'c}}, {Jester}, {Jiang}, {Johnson}, {Jorgensen}, {Juri{\'c}}, {Kent}, {Kessler}, {Kleinman}, {Knapp},
  {Konishi}, {Kron}, {Krzesinski}, {Kuropatkin}, {Lampeitl}, {Lebedeva}, {Lee}, {Lee}, {French Leger}, {L{\'e}pine}, {Li}, {Lima}, {Lin}, {Long}, {Loomis}, {Loveday}, {Lupton}, {Magnier}, {Malanushenko}, {Malanushenko}, {Mandelbaum}, {Margon}, {Marriner}, {Mart{\'\i}nez-Delgado}, {Matsubara}, {McGehee}, {McKay}, {Meiksin}, {Morrison}, {Mullally}, {Munn}, {Murphy}, {Nash}, {Nebot}, {Neilsen}, {Newberg}, {Newman}, {Nichol}, {Nicinski}, {Nieto-Santisteban}, {Nitta}, {Okamura}, {Oravetz}, {Ostriker}, {Owen}, {Padmanabhan}, {Pan}, {Park}, {Pauls}, {Peoples}, {Percival}, {Pier}, {Pope}, {Pourbaix}, {Price}, {Purger}, {Quinn}, {Raddick}, {Re Fiorentin}, {Richards}, {Richmond}, {Riess}, {Rix}, {Rockosi}, {Sako}, {Schlegel}, {Schneider}, {Scholz}, {Schreiber}, {Schwope}, {Seljak}, {Sesar}, {Sheldon}, {Shimasaku}, {Sibley}, {Simmons}, {Sivarani}, {Allyn Smith}, {Smith}, {Smol{\v{c}}i{\'c}}, {Snedden}, {Stebbins}, {Steinmetz}, {Stoughton}, {Strauss}, {SubbaRao}, {Suto}, {Szalay}, {Szapudi}, {Szkody}, {Tanaka},
  {Tegmark}, {Teodoro}, {Thakar}, {Tremonti}, {Tucker}, {Uomoto}, {Vanden Berk}, {Vandenberg}, {Vidrih}, {Vogeley}, {Voges}, {Vogt}, {Wadadekar}, {Watters}, {Weinberg}, {West}, {White}, {Wilhite}, {Wonders}, {Yanny}, \& {Yocum}}]{SDSS}
{Abazajian}, K.~N., {Adelman-McCarthy}, J.~K., {Ag{\"u}eros}, M.~A., {et~al.} 2009, \apjs, 182, 543

\bibitem[{{Arias} {et~al.}(2025){Arias}, {Bell}, {Gozman}, {Jang}, {Stockton}, {Gnedin}, {D'Souza}, {Monachesi}, {Bailin}, {Nidever}, \& {de Jong}}]{And35}
{Arias}, J.~M., {Bell}, E.~F., {Gozman}, K., {et~al.} 2025, \apjl, 982, L3

\bibitem[{{Beasley} {et~al.}(2019){Beasley}, {Leaman}, {Gallart}, {Larsen}, {Battaglia}, {Monelli}, \& {Pedreros}}]{Beasley2019}
{Beasley}, M.~A., {Leaman}, R., {Gallart}, C., {et~al.} 2019, \mnras, 487, 1986

\bibitem[{{Belokurov} {et~al.}(2006){Belokurov}, {Zucker}, {Evans}, {Wilkinson}, {Irwin}, {Hodgkin}, {Bramich}, {Irwin}, {Gilmore}, {Willman}, {Vidrih}, {Newberg}, {Wyse}, {Fellhauer}, {Hewett}, {Cole}, {Bell}, {Beers}, {Rockosi}, {Yanny}, {Grebel}, {Schneider}, {Lupton}, {Barentine}, {Brewington}, {Brinkmann}, {Harvanek}, {Kleinman}, {Krzesinski}, {Long}, {Nitta}, {Smith}, \& {Snedden}}]{Belokurov2006}
{Belokurov}, V., {Zucker}, D.~B., {Evans}, N.~W., {et~al.} 2006, \apjl, 647, L111

\bibitem[{{Bressan} {et~al.}(2012){Bressan}, {Marigo}, {Girardi}, {Salasnich}, {Dal Cero}, {Rubele}, \& {Nanni}}]{Bressan2012}
{Bressan}, A., {Marigo}, P., {Girardi}, L., {et~al.} 2012, \mnras, 427, 127

\bibitem[{{Bullock} \& {Boylan-Kolchin}(2017)}]{Bullock2017}
{Bullock}, J.~S. \& {Boylan-Kolchin}, M. 2017, \araa, 55, 343

\bibitem[{{{\c{C}}al{\i}{\textcommabelow s}kan} {et~al.}(2012){{\c{C}}al{\i}{\textcommabelow s}kan}, {Christlieb}, \& {Grebel}}]{Caliskan2012}
{{\c{C}}al{\i}{\textcommabelow s}kan}, {\c{S}}., {Christlieb}, N., \& {Grebel}, E.~K. 2012, \aap, 537, A83

\bibitem[{{Cerny} {et~al.}(2026){Cerny}, {Geha}, {Simon}, {Li}, {Pace}, {van der Marel}, {Kallivayalil}, {Ji}, {Drlica-Wagner}, {Crnojevi{\'c}}, {Limberg}, {No{\"e}l}, {Riley}, {Sand}, {Stringfellow}, {Choi}, {Ferguson}, {Mart{\'\i}nez-V{\'a}zquez}, {Massana}, {Medina}, {Mutlu-Pakdil}, {Sakowska}, {Tan}, {Vivas}, \& {Bruce}}]{Cerny2026}
{Cerny}, W., {Geha}, M., {Simon}, J.~D., {et~al.} 2026, arXiv e-prints, arXiv:2610.07182

\bibitem[{{Chambers} {et~al.}(2016){Chambers}, {Magnier}, {Metcalfe}, {Flewelling}, {Huber}, {Waters}, {Denneau}, {Draper}, {Farrow}, {Finkbeiner}, {Holmberg}, {Koppenhoefer}, {Price}, {Rest}, {Saglia}, {Schlafly}, {Smartt}, {Sweeney}, {Wainscoat}, {Burgett}, {Chastel}, {Grav}, {Heasley}, {Hodapp}, {Jedicke}, {Kaiser}, {Kudritzki}, {Luppino}, {Lupton}, {Monet}, {Morgan}, {Onaka}, {Shiao}, {Stubbs}, {Tonry}, {White}, {Ba{\~n}ados}, {Bell}, {Bender}, {Bernard}, {Boegner}, {Boffi}, {Botticella}, {Calamida}, {Casertano}, {Chen}, {Chen}, {Cole}, {Deacon}, {Frenk}, {Fitzsimmons}, {Gezari}, {Gibbs}, {Goessl}, {Goggia}, {Gourgue}, {Goldman}, {Grant}, {Grebel}, {Hambly}, {Hasinger}, {Heavens}, {Heckman}, {Henderson}, {Henning}, {Holman}, {Hopp}, {Ip}, {Isani}, {Jackson}, {Keyes}, {Koekemoer}, {Kotak}, {Le}, {Liska}, {Long}, {Lucey}, {Liu}, {Martin}, {Masci}, {McLean}, {Mindel}, {Misra}, {Morganson}, {Murphy}, {Obaika}, {Narayan}, {Nieto-Santisteban}, {Norberg}, {Peacock}, {Pier}, {Postman}, {Primak}, {Rae}, {Rai},
  {Riess}, {Riffeser}, {Rix}, {R{\"o}ser}, {Russel}, {Rutz}, {Schilbach}, {Schultz}, {Scolnic}, {Strolger}, {Szalay}, {Seitz}, {Small}, {Smith}, {Soderblom}, {Taylor}, {Thomson}, {Taylor}, {Thakar}, {Thiel}, {Thilker}, {Unger}, {Urata}, {Valenti}, {Wagner}, {Walder}, {Walter}, {Watters}, {Werner}, {Wood-Vasey}, \& {Wyse}}]{Chambers2016}
{Chambers}, K.~C., {Magnier}, E.~A., {Metcalfe}, N., {et~al.} 2016, arXiv e-prints, arXiv:1612.05560

\bibitem[{{Chapman} {et~al.}(2008){Chapman}, {Ibata}, {Irwin}, {Koch}, {Letarte}, {Martin}, {Collins}, {Lewis}, {McConnachie}, {Pe{\~n}arrubia}, {Rich}, {Trethewey}, {Ferguson}, {Huxor}, \& {Tanvir}}]{Chapman2008}
{Chapman}, S.~C., {Ibata}, R., {Irwin}, M., {et~al.} 2008, \mnras, 390, 1437

\bibitem[{{Collins} {et~al.}(2009){Collins}, {Chapman}, {Irwin}, {Ibata}, {Martin}, {Ferguson}, {Huxor}, {Lewis}, {Mackey}, {McConnachie}, \& {Tanvir}}]{Collins2009}
{Collins}, M.~L.~M., {Chapman}, S.~C., {Irwin}, M., {et~al.} 2009, \mnras, 396, 1619

\bibitem[{{Collins} {et~al.}(2022){Collins}, {Charles}, {Mart{\'\i}nez-Delgado}, {Monelli}, {Karim}, {Donatiello}, {Tollerud}, \& {Boschin}}]{Collins2022}
{Collins}, M. L.~M., {Charles}, E. J.~E., {Mart{\'\i}nez-Delgado}, D., {et~al.} 2022, \mnras, 515, L72

\bibitem[{{Collins} {et~al.}(2024){Collins}, {Karim}, {Martinez-Delgado}, {Monelli}, {Tollerud}, {Donatiello}, {Navabi}, {Charles}, \& {Boschin}}]{Collins2024}
{Collins}, M. L.~M., {Karim}, N., {Martinez-Delgado}, D., {et~al.} 2024, \mnras, 528, 2614

\bibitem[{{DES Collaboration} {et~al.}(2016){DES Collaboration}, {Abbott}, {Abdalla}, {Aleksi{\'c}}, {Allam}, {Amara}, {Bacon}, {Balbinot}, {Banerji}, {Bechtol}, {Benoit-L{\'e}vy}, {Bernstein}, {Bertin}, {Blazek}, {Bonnett}, {Bridle}, {Brooks}, {Brunner}, {Buckley-Geer}, {Burke}, {Caminha}, {Capozzi}, {Carlsen}, {Carnero-Rosell}, {Carollo}, {Carrasco-Kind}, {Carretero}, {Castander}, {Clerkin}, {Collett}, {Conselice}, {Crocce}, {Cunha}, {D'Andrea}, {da Costa}, {Davis}, {Desai}, {Diehl}, {Dietrich}, {Dodelson}, {Doel}, {Drlica-Wagner}, {Estrada}, {Etherington}, {Evrard}, {Fabbri}, {Finley}, {Flaugher}, {Foley}, {Fosalba}, {Frieman}, {Garc{\'\i}a-Bellido}, {Gaztanaga}, {Gerdes}, {Giannantonio}, {Goldstein}, {Gruen}, {Gruendl}, {Guarnieri}, {Gutierrez}, {Hartley}, {Honscheid}, {Jain}, {James}, {Jeltema}, {Jouvel}, {Kessler}, {King}, {Kirk}, {Kron}, {Kuehn}, {Kuropatkin}, {Lahav}, {Li}, {Lima}, {Lin}, {Maia}, {Makler}, {Manera}, {Maraston}, {Marshall}, {Martini}, {McMahon}, {Melchior}, {Merson}, {Miller},
  {Miquel}, {Mohr}, {Morice-Atkinson}, {Naidoo}, {Neilsen}, {Nichol}, {Nord}, {Ogando}, {Ostrovski}, {Palmese}, {Papadopoulos}, {Peiris}, {Peoples}, {Percival}, {Plazas}, {Reed}, {Refregier}, {Romer}, {Roodman}, {Ross}, {Rozo}, {Rykoff}, {Sadeh}, {Sako}, {S{\'a}nchez}, {Sanchez}, {Santiago}, {Scarpine}, {Schubnell}, {Sevilla-Noarbe}, {Sheldon}, {Smith}, {Smith}, {Soares-Santos}, {Sobreira}, {Soumagnac}, {Suchyta}, {Sullivan}, {Swanson}, {Tarle}, {Thaler}, {Thomas}, {Thomas}, {Tucker}, {Vieira}, {Vikram}, {Walker}, {Wechsler}, {Weller}, {Wester}, {Whiteway}, {Wilcox}, {Yanny}, {Zhang}, \& {Zuntz}}]{DES}
{DES Collaboration}, {Abbott}, T., {Abdalla}, F.~B., {et~al.} 2016, \mnras, 460, 1270

\bibitem[{{Dey} {et~al.}(2019){Dey}, {Schlegel}, {Lang}, {Blum}, {Burleigh}, {Fan}, {Findlay}, {Finkbeiner}, {Herrera}, {Juneau}, {Landriau}, {Levi}, {McGreer}, {Meisner}, {Myers}, {Moustakas}, {Nugent}, {Patej}, {Schlafly}, {Walker}, {Valdes}, {Weaver}, {Y{\`e}che}, {Zou}, {Zhou}, {Abareshi}, {Abbott}, {Abolfathi}, {Aguilera}, {Alam}, {Allen}, {Alvarez}, {Annis}, {Ansarinejad}, {Aubert}, {Beechert}, {Bell}, {BenZvi}, {Beutler}, {Bielby}, {Bolton}, {Brice{\~n}o}, {Buckley-Geer}, {Butler}, {Calamida}, {Carlberg}, {Carter}, {Casas}, {Castander}, {Choi}, {Comparat}, {Cukanovaite}, {Delubac}, {DeVries}, {Dey}, {Dhungana}, {Dickinson}, {Ding}, {Donaldson}, {Duan}, {Duckworth}, {Eftekharzadeh}, {Eisenstein}, {Etourneau}, {Fagrelius}, {Farihi}, {Fitzpatrick}, {Font-Ribera}, {Fulmer}, {G{\"a}nsicke}, {Gaztanaga}, {George}, {Gerdes}, {Gontcho}, {Gorgoni}, {Green}, {Guy}, {Harmer}, {Hernandez}, {Honscheid}, {Huang}, {James}, {Jannuzi}, {Jiang}, {Joyce}, {Karcher}, {Karkar}, {Kehoe}, {Kneib}, {Kueter-Young}, {Lan},
  {Lauer}, {Le Guillou}, {Le Van Suu}, {Lee}, {Lesser}, {Perreault Levasseur}, {Li}, {Mann}, {Marshall}, {Mart{\'\i}nez-V{\'a}zquez}, {Martini}, {du Mas des Bourboux}, {McManus}, {Meier}, {M{\'e}nard}, {Metcalfe}, {Mu{\~n}oz-Guti{\'e}rrez}, {Najita}, {Napier}, {Narayan}, {Newman}, {Nie}, {Nord}, {Norman}, {Olsen}, {Paat}, {Palanque-Delabrouille}, {Peng}, {Poppett}, {Poremba}, {Prakash}, {Rabinowitz}, {Raichoor}, {Rezaie}, {Robertson}, {Roe}, {Ross}, {Ross}, {Rudnick}, {Safonova}, {Saha}, {S{\'a}nchez}, {Savary}, {Schweiker}, {Scott}, {Seo}, {Shan}, {Silva}, {Slepian}, {Soto}, {Sprayberry}, {Staten}, {Stillman}, {Stupak}, {Summers}, {Sien Tie}, {Tirado}, {Vargas-Maga{\~n}a}, {Vivas}, {Wechsler}, {Williams}, {Yang}, {Yang}, {Yapici}, {Zaritsky}, {Zenteno}, {Zhang}, {Zhang}, {Zhou}, \& {Zhou}}]{Dey2019}
{Dey}, A., {Schlegel}, D.~J., {Lang}, D., {et~al.} 2019, \aj, 157, 168

\bibitem[{Donatiello(2025)}]{Donatiello2025}
Donatiello, G. 2025, Andromeda 450 kpc Survey - Deep survey of the halo of M31 - Satellite galaxies map

\bibitem[{{Drlica-Wagner} {et~al.}(2021){Drlica-Wagner}, {Carlin}, {Nidever}, {Ferguson}, {Kuropatkin}, {Adam{\'o}w}, {Cerny}, {Choi}, {Esteves}, {Mart{\'\i}nez-V{\'a}zquez}, {Mau}, {Miller}, {Mutlu-Pakdil}, {Neilsen}, {Olsen}, {Pace}, {Riley}, {Sakowska}, {Sand}, {Santana-Silva}, {Tollerud}, {Tucker}, {Vivas}, {Zaborowski}, {Zenteno}, {Abbott}, {Allam}, {Bechtol}, {Bell}, {Bell}, {Bilaji}, {Bom}, {Carballo-Bello}, {Crnojevi{\'c}}, {Cioni}, {Diaz-Ocampo}, {de Boer}, {Erkal}, {Gruendl}, {Hernandez-Lang}, {Hughes}, {James}, {Johnson}, {Li}, {Mao}, {Mart{\'\i}nez-Delgado}, {Massana}, {McNanna}, {Morgan}, {Nadler}, {No{\"e}l}, {Palmese}, {Peter}, {Rykoff}, {S{\'a}nchez}, {Shipp}, {Simon}, {Smercina}, {Soares-Santos}, {Stringfellow}, {Tavangar}, {van der Marel}, {Walker}, {Wechsler}, {Wu}, {Yanny}, {Fitzpatrick}, {Huang}, {Jacques}, {Nikutta}, {Scott}, \& {Astro Data Lab}}]{Drlica-Wagner2021}
{Drlica-Wagner}, A., {Carlin}, J.~L., {Nidever}, D.~L., {et~al.} 2021, \apjs, 256, 2

\bibitem[{{Drlica-Wagner} {et~al.}(2022){Drlica-Wagner}, {Ferguson}, {Adam{\'o}w}, {Aguena}, {Allam}, {Andrade-Oliveira}, {Bacon}, {Bechtol}, {Bell}, {Bertin}, {Bilaji}, {Bocquet}, {Bom}, {Brooks}, {Burke}, {Carballo-Bello}, {Carlin}, {Carnero Rosell}, {Carrasco Kind}, {Carretero}, {Castander}, {Cerny}, {Chang}, {Choi}, {Conselice}, {Costanzi}, {Crnojevi{\'c}}, {da Costa}, {de Vicente}, {Desai}, {Esteves}, {Everett}, {Ferrero}, {Fitzpatrick}, {Flaugher}, {Friedel}, {Frieman}, {Garc{\'\i}a-Bellido}, {Gatti}, {Gaztanaga}, {Gerdes}, {Gruen}, {Gruendl}, {Gschwend}, {Hartley}, {Hernandez-Lang}, {Hinton}, {Hollowood}, {Honscheid}, {Hughes}, {Jacques}, {James}, {Johnson}, {Kuehn}, {Kuropatkin}, {Lahav}, {Li}, {Lidman}, {Lin}, {March}, {Marshall}, {Mart{\'\i}nez-Delgado}, {Mart{\'\i}nez-V{\'a}zquez}, {Massana}, {Mau}, {McNanna}, {Melchior}, {Menanteau}, {Miller}, {Miquel}, {Mohr}, {Morgan}, {Mutlu-Pakdil}, {Mu{\~n}oz}, {Neilsen}, {Nidever}, {Nikutta}, {Nilo Castellon}, {No{\"e}l}, {Ogando}, {Olsen}, {Pace},
  {Palmese}, {Paz-Chinch{\'o}n}, {Pereira}, {Pieres}, {Plazas Malag{\'o}n}, {Prat}, {Riley}, {Rodriguez-Monroy}, {Romer}, {Roodman}, {Sako}, {Sakowska}, {Sanchez}, {S{\'a}nchez}, {Sand}, {Santana-Silva}, {Santiago}, {Schubnell}, {Serrano}, {Sevilla-Noarbe}, {Simon}, {Smith}, {Soares-Santos}, {Stringfellow}, {Suchyta}, {Suson}, {Tan}, {Tarle}, {Tavangar}, {Thomas}, {To}, {Tollerud}, {Troxel}, {Tucker}, {Varga}, {Vivas}, {Walker}, {Weller}, {Wilkinson}, {Wu}, {Yanny}, {Zaborowski}, {Zenteno}, {Delve Collaboration}, {Des Collaboration}, \& {Astro Data Lab}}]{Drlica-Wagner2022}
{Drlica-Wagner}, A., {Ferguson}, P.~S., {Adam{\'o}w}, M., {et~al.} 2022, \apjs, 261, 38

\bibitem[{{Euclid Collaboration} {et~al.}(2026){Euclid Collaboration}, {Gatto}, {Rusterucci}, {Bellazzini}, {Martin}, {Ferguson}, {Libralato}, {Pascale}, {Bedin}, {Saifollahi}, \& et~al.}]{Gatto2026}
{Euclid Collaboration}, {Gatto}, M., {Rusterucci}, S., {et~al.} 2026, arXiv e-prints, arXiv:2611.16025

\bibitem[{{Forbes} \& {Bridges}(2010)}]{Forbes2010}
{Forbes}, D.~A. \& {Bridges}, T. 2010, \mnras, 404, 1203

\bibitem[{{Foreman-Mackey} {et~al.}(2013){Foreman-Mackey}, {Hogg}, {Lang}, \& {Goodman}}]{Foreman-Mackey2013}
{Foreman-Mackey}, D., {Hogg}, D.~W., {Lang}, D., \& {Goodman}, J. 2013, \pasp, 125, 306

\bibitem[{{Gallart} {et~al.}(2005){Gallart}, {Zoccali}, \& {Aparicio}}]{Gallart2005}
{Gallart}, C., {Zoccali}, M., \& {Aparicio}, A. 2005, \araa, 43, 387

\bibitem[{{Gatto} {et~al.}(2024){Gatto}, {Bellazzini}, {Tortora}, {Ripepi}, {Dall'Ora}, {Cignoni}, {Kuijken}, {Hildebrandt}, {Zhang}, {de Jong}, {Napolitano}, \& {Smith}}]{Gatto2024}
{Gatto}, M., {Bellazzini}, M., {Tortora}, C., {et~al.} 2024, \aap, 681, L13

\bibitem[{{Gnedin} \& {Kravtsov}(2006)}]{Gnedin2006}
{Gnedin}, N.~Y. \& {Kravtsov}, A.~V. 2006, \apj, 645, 1054

\bibitem[{{Grebel}(1997)}]{Grebel1997}
{Grebel}, E.~K. 1997, Reviews in Modern Astronomy, 10, 29

\bibitem[{{Gwyn} {et~al.}(2025){Gwyn}, {McConnachie}, {Cuillandre}, {Chambers}, {Magnier}, {de Boer}, {Hudson}, {Oguri}, {Furusawa}, {Hildebrandt}, {Carlberg}, {Ellison}, {Furusawa}, {Gavazzi}, {Ibata}, {Mellier}, {Osato}, {Aussel}, {Baumont}, {Bayer}, {Boulade}, {C{\^o}t{\'e}}, {Chemaly}, {Daley}, {Duc}, {Durret}, {Ellien}, {Fabbro}, {Ferreira}, {Fitriana}, {Le Floc'h}, {Fudamoto}, {Gao}, {Goh}, {Goto}, {Guerrini}, {Guinot}, {H{\'e}nault-Brunet}, {Hammer}, {Harikane}, {Hayashi}, {Heesters}, {Ichikawa}, {Kilbinger}, {Kuzma}, {Li}, {Liaudat}, {Lin}, {M{\"u}ller}, {Martin}, {Matsuoka}, {Medina}, {Miyatake}, {Miyazaki}, {Mpetha}, {Nagao}, {Navarro}, {Niwano}, {Ogami}, {Okabe}, {Onoue}, {Paek}, {Parker}, {Patton}, {Peters}, {Prunet}, {S{\'a}nchez-Janssen}, {Schultheis}, {Sestito}, {Smith}, {Starck}, {Starkenburg}, {Stone}, {Storfer}, {Suzuki}, {Erben}, {Taibi}, {Thomas}, {Toba}, {Uchiyama}, {Valls-Gabaud}, {Venn}, {Van Waerbeke}, {Wainscoat}, {Wilkinson}, {Wittje}, {Yoshida}, {Zhang}, \& {Zhong}}]{unions}
{Gwyn}, S., {McConnachie}, A.~W., {Cuillandre}, J.-C., {et~al.} 2025, \aj, 170, 324

\bibitem[{{Huxor} {et~al.}(2011){Huxor}, {Ferguson}, {Tanvir}, {Irwin}, {Mackey}, {Ibata}, {Bridges}, {Chapman}, \& {Lewis}}]{Huxor2011}
{Huxor}, A.~P., {Ferguson}, A.~M.~N., {Tanvir}, N.~R., {et~al.} 2011, \mnras, 414, 770

\bibitem[{{Huxor} {et~al.}(2014){Huxor}, {Mackey}, {Ferguson}, {Irwin}, {Martin}, {Tanvir}, {Veljanoski}, {McConnachie}, {Fishlock}, {Ibata}, \& {Lewis}}]{Huxor2014}
{Huxor}, A.~P., {Mackey}, A.~D., {Ferguson}, A.~M.~N., {et~al.} 2014, \mnras, 442, 2165

\bibitem[{{Huxor} {et~al.}(2005){Huxor}, {Tanvir}, {Irwin}, {Ibata}, {Collett}, {Ferguson}, {Bridges}, \& {Lewis}}]{Huxor2005}
{Huxor}, A.~P., {Tanvir}, N.~R., {Irwin}, M.~J., {et~al.} 2005, \mnras, 360, 1007

\bibitem[{{Ibata} {et~al.}(2007){Ibata}, {Martin}, {Irwin}, {Chapman}, {Ferguson}, {Lewis}, \& {McConnachie}}]{ibata2007}
{Ibata}, R., {Martin}, N.~F., {Irwin}, M., {et~al.} 2007, \apj, 671, 1591

\bibitem[{{Ibata} {et~al.}(2014){Ibata}, {Lewis}, {McConnachie}, {Martin}, {Irwin}, {Ferguson}, {Babul}, {Bernard}, {Chapman}, {Collins}, {Fardal}, {Mackey}, {Navarro}, {Pe{\~n}arrubia}, {Rich}, {Tanvir}, \& {Widrow}}]{Ibata2014}
{Ibata}, R.~A., {Lewis}, G.~F., {McConnachie}, A.~W., {et~al.} 2014, \apj, 780, 128

\bibitem[{{Illingworth}(1976)}]{Illingworth1976}
{Illingworth}, G. 1976, \apj, 204, 73

\bibitem[{{Jordi} {et~al.}(2006){Jordi}, {Grebel}, \& {Ammon}}]{Jordi2006}
{Jordi}, K., {Grebel}, E.~K., \& {Ammon}, K. 2006, \aap, 460, 339

\bibitem[{{Kirby} {et~al.}(2013){Kirby}, {Cohen}, {Guhathakurta}, {Cheng}, {Bullock}, \& {Gallazzi}}]{Kirby2013}
{Kirby}, E.~N., {Cohen}, J.~G., {Guhathakurta}, P., {et~al.} 2013, \apj, 779, 102

\bibitem[{{Kirby} {et~al.}(2011){Kirby}, {Lanfranchi}, {Simon}, {Cohen}, \& {Guhathakurta}}]{Kirby2011}
{Kirby}, E.~N., {Lanfranchi}, G.~A., {Simon}, J.~D., {Cohen}, J.~G., \& {Guhathakurta}, P. 2011, \apj, 727, 78

\bibitem[{{Kraft}(1994)}]{Kraft1994}
{Kraft}, R.~P. 1994, \pasp, 106, 553

\bibitem[{{Kroupa}(2002)}]{Kroupa2002}
{Kroupa}, P. 2002, Science, 295, 82

\bibitem[{{Kruijssen}(2019)}]{Kruijssen2019}
{Kruijssen}, J.~M.~D. 2019, \mnras, 486, L20

\bibitem[{{Laevens} {et~al.}(2015){Laevens}, {Martin}, {Bernard}, {Schlafly}, {Sesar}, {Rix}, {Bell}, {Ferguson}, {Slater}, {Sweeney}, {Wyse}, {Huxor}, {Burgett}, {Chambers}, {Draper}, {Hodapp}, {Kaiser}, {Magnier}, {Metcalfe}, {Tonry}, {Wainscoat}, \& {Waters}}]{Laevens2015}
{Laevens}, B. P.~M., {Martin}, N.~F., {Bernard}, E.~J., {et~al.} 2015, \apj, 813, 44

\bibitem[{{Longeard} {et~al.}(2021){Longeard}, {Martin}, {Ibata}, {Starkenburg}, {Jablonka}, {Aguado}, {Carlberg}, {C{\^o}t{\'e}}, {Gonz{\'a}lez Hern{\'a}ndez}, {Lucchesi}, {Malhan}, {Navarro}, {S{\'a}nchez-Janssen}, {Thomas}, {Venn}, \& {McConnachie}}]{Longeard2021}
{Longeard}, N., {Martin}, N., {Ibata}, R.~A., {et~al.} 2021, \mnras, 503, 2754

\bibitem[{{Mackey} {et~al.}(2019){Mackey}, {Ferguson}, {Huxor}, {Veljanoski}, {Lewis}, {McConnachie}, {Martin}, {Ibata}, {Irwin}, {C{\^o}t{\'e}}, {Collins}, {Tanvir}, \& {Bate}}]{Mackey2019}
{Mackey}, A.~D., {Ferguson}, A.~M.~N., {Huxor}, A.~P., {et~al.} 2019, \mnras, 484, 1756

\bibitem[{{Mackey} {et~al.}(2006){Mackey}, {Huxor}, {Ferguson}, {Tanvir}, {Irwin}, {Ibata}, {Bridges}, {Johnson}, \& {Lewis}}]{Mackey2006}
{Mackey}, A.~D., {Huxor}, A., {Ferguson}, A.~M.~N., {et~al.} 2006, \apjl, 653, L105

\bibitem[{{Mackey} {et~al.}(2010){Mackey}, {Huxor}, {Ferguson}, {Irwin}, {Tanvir}, {McConnachie}, {Ibata}, {Chapman}, \& {Lewis}}]{Mackey2010}
{Mackey}, A.~D., {Huxor}, A.~P., {Ferguson}, A.~M.~N., {et~al.} 2010, \apjl, 717, L11

\bibitem[{{Mackey} {et~al.}(2013){Mackey}, {Huxor}, {Martin}, {Ferguson}, {Dotter}, {McConnachie}, {Ibata}, {Irwin}, {Lewis}, {Sakari}, {Tanvir}, \& {Venn}}]{Mackey2013b}
{Mackey}, A.~D., {Huxor}, A.~P., {Martin}, N.~F., {et~al.} 2013, \apjl, 770, L17

\bibitem[{{Martin} {et~al.}(2008){Martin}, {de Jong}, \& {Rix}}]{Martin2008}
{Martin}, N.~F., {de Jong}, J. T.~A., \& {Rix}, H.-W. 2008, \apj, 684, 1075

\bibitem[{{Martin} {et~al.}(2006){Martin}, {Ibata}, {Irwin}, {Chapman}, {Lewis}, {Ferguson}, {Tanvir}, \& {McConnachie}}]{martin2006}
{Martin}, N.~F., {Ibata}, R.~A., {Irwin}, M.~J., {et~al.} 2006, \mnras, 371, 1983

\bibitem[{{Martin} {et~al.}(2016){Martin}, {Ibata}, {Lewis}, {McConnachie}, {Babul}, {Bate}, {Bernard}, {Chapman}, {Collins}, {Conn}, {Crnojevi{\'c}}, {Fardal}, {Ferguson}, {Irwin}, {Mackey}, {McMonigal}, {Navarro}, \& {Rich}}]{Martin2016b}
{Martin}, N.~F., {Ibata}, R.~A., {Lewis}, G.~F., {et~al.} 2016, \apj, 833, 167

\bibitem[{{Martin} {et~al.}(2009){Martin}, {McConnachie}, {Irwin}, {Widrow}, {Ferguson}, {Ibata}, {Dubinski}, {Babul}, {Chapman}, {Fardal}, {Lewis}, {Navarro}, \& {Rich}}]{Martin2009}
{Martin}, N.~F., {McConnachie}, A.~W., {Irwin}, M., {et~al.} 2009, \apj, 705, 758

\bibitem[{{Mart{\'\i}nez-Delgado} {et~al.}(2018){Mart{\'\i}nez-Delgado}, {Grebel}, {Javanmardi}, {Boschin}, {Longeard}, {Carballo-Bello}, {Makarov}, {Beasley}, {Donatiello}, {Haynes}, {Forbes}, \& {Romanowsky}}]{MD2018}
{Mart{\'\i}nez-Delgado}, D., {Grebel}, E.~K., {Javanmardi}, B., {et~al.} 2018, \aap, 620, A126

\bibitem[{{Mart{\'\i}nez-Delgado} {et~al.}(2022){Mart{\'\i}nez-Delgado}, {Karim}, {Charles}, {Boschin}, {Monelli}, {Collins}, {Donatiello}, \& {Alfaro}}]{MD2022}
{Mart{\'\i}nez-Delgado}, D., {Karim}, N., {Charles}, E. J.~E., {et~al.} 2022, \mnras, 509, 16

\bibitem[{{Mart{\'\i}nez-Delgado} {et~al.}(2021){Mart{\'\i}nez-Delgado}, {Makarov}, {Javanmardi}, {Pawlowski}, {Makarova}, {Donatiello}, {Lang}, {Rom{\'a}n}, {Vivas}, \& {Carballo-Bello}}]{MD2021}
{Mart{\'\i}nez-Delgado}, D., {Makarov}, D., {Javanmardi}, B., {et~al.} 2021, \aap, 652, A48

\bibitem[{{Mart{\'\i}nez-Delgado} {et~al.}(2026){Mart{\'\i}nez-Delgado}, {Stein}, {Pawlowski}, {Sakowska}, {Makarov}, {Makarova}, {Donatiello}, \& {Lang}}]{MD2025}
{Mart{\'\i}nez-Delgado}, D., {Stein}, M., {Pawlowski}, M.~S., {et~al.} 2026, \aap, 705, A77

\bibitem[{{McConnachie} {et~al.}(2008){McConnachie}, {Huxor}, {Martin}, {Irwin}, {Chapman}, {Fahlman}, {Ferguson}, {Ibata}, {Lewis}, {Richer}, \& {Tanvir}}]{McConnachie2008}
{McConnachie}, A.~W., {Huxor}, A., {Martin}, N.~F., {et~al.} 2008, \apj, 688, 1009

\bibitem[{{McConnachie} {et~al.}(2018){McConnachie}, {Ibata}, {Martin}, {Ferguson}, {Collins}, {Gwyn}, {Irwin}, {Lewis}, {Mackey}, {Davidge}, {Arias}, {Conn}, {C{\^o}t{\'e}}, {Crnojevic}, {Huxor}, {Penarrubia}, {Spengler}, {Tanvir}, {Valls-Gabaud}, {Babul}, {Barmby}, {Bate}, {Bernard}, {Chapman}, {Dotter}, {Harris}, {McMonigal}, {Navarro}, {Puzia}, {Rich}, {Thomas}, \& {Widrow}}]{McConnachie2018}
{McConnachie}, A.~W., {Ibata}, R., {Martin}, N., {et~al.} 2018, \apj, 868, 55

\bibitem[{{McConnachie} {et~al.}(2009){McConnachie}, {Irwin}, {Ibata}, {Dubinski}, {Widrow}, {Martin}, {C{\^o}t{\'e}}, {Dotter}, {Navarro}, {Ferguson}, {Puzia}, {Lewis}, {Babul}, {Barmby}, {Bienaym{\'e}}, {Chapman}, {Cockcroft}, {Collins}, {Fardal}, {Harris}, {Huxor}, {Mackey}, {Pe{\~n}arrubia}, {Rich}, {Richer}, {Siebert}, {Tanvir}, {Valls-Gabaud}, \& {Venn}}]{McConnachie2009}
{McConnachie}, A.~W., {Irwin}, M.~J., {Ibata}, R.~A., {et~al.} 2009, \nat, 461, 66

\bibitem[{{McGill} {et~al.}(2025){McGill}, {Ferguson}, {Mackey}, {Huxor}, {Lewis}, {Martin}, {McConnachie}, {Sakari}, {Tanvir}, \& {Venn}}]{McGill2025}
{McGill}, G., {Ferguson}, A. M.~N., {Mackey}, D., {et~al.} 2025, \mnras, 542, L60

\bibitem[{{Monelli} {et~al.}(2010){Monelli}, {Hidalgo}, {Stetson}, {Aparicio}, {Gallart}, {Dolphin}, {Cole}, {Weisz}, {Skillman}, {Bernard}, {Mayer}, {Navarro}, {Cassisi}, {Drozdovsky}, \& {Tolstoy}}]{Monelli2010}
{Monelli}, M., {Hidalgo}, S.~L., {Stetson}, P.~B., {et~al.} 2010, \apj, 720, 1225

\bibitem[{{Pace}(2025)}]{Pace2025}
{Pace}, A.~B. 2025, The Open Journal of Astrophysics, 8, 142

\bibitem[{{Peterson} \& {Latham}(1986)}]{Peterson1986}
{Peterson}, R.~C. \& {Latham}, D.~W. 1986, \apj, 305, 645

\bibitem[{{Richardson} {et~al.}(2011){Richardson}, {Irwin}, {McConnachie}, {Martin}, {Dotter}, {Ferguson}, {Ibata}, {Chapman}, {Lewis}, {Tanvir}, \& {Rich}}]{Richardson2011}
{Richardson}, J.~C., {Irwin}, M.~J., {McConnachie}, A.~W., {et~al.} 2011, \apj, 732, 76

\bibitem[{{Sakowska} {et~al.}(2026){Sakowska}, {Mart{\'\i}nez-Delgado}, {Collins}, {Monelli}, {Donatiello}, {Doliva-Dolinsky}, \& {Santos-Santos}}]{Sakowska2026}
{Sakowska}, J.~D., {Mart{\'\i}nez-Delgado}, D., {Collins}, M. L.~M., {et~al.} 2026, \aap, 710, A342

\bibitem[{{Savino} {et~al.}(2022){Savino}, {Weisz}, {Skillman}, {Dolphin}, {Kallivayalil}, {Wetzel}, {Anderson}, {Besla}, {Boylan-Kolchin}, {Bullock}, {Cole}, {Collins}, {Cooper}, {Deason}, {Dotter}, {Fardal}, {Ferguson}, {Fritz}, {Geha}, {Gilbert}, {Guhathakurta}, {Ibata}, {Irwin}, {Jeon}, {Kirby}, {Lewis}, {Mackey}, {Majewski}, {Martin}, {McConnachie}, {Patel}, {Rich}, {Simon}, {Sohn}, {Tollerud}, \& {van der Marel}}]{Savino2022}
{Savino}, A., {Weisz}, D.~R., {Skillman}, E.~D., {et~al.} 2022, \apj, 938, 101

\bibitem[{{Schlafly} \& {Finkbeiner}(2011)}]{Schlafly2011}
{Schlafly}, E.~F. \& {Finkbeiner}, D.~P. 2011, \apj, 737, 103

\bibitem[{{Schlegel} {et~al.}(1998){Schlegel}, {Finkbeiner}, \& {Davis}}]{Schlegel1998}
{Schlegel}, D.~J., {Finkbeiner}, D.~P., \& {Davis}, M. 1998, \apj, 500, 525

\bibitem[{{Simon}(2019)}]{Simon2019}
{Simon}, J.~D. 2019, \araa, 57, 375

\bibitem[{{Simon} \& {Geha}(2007)}]{SimonGeha2007}
{Simon}, J.~D. \& {Geha}, M. 2007, \apj, 670, 313

\bibitem[{{Smith} {et~al.}(2025){Smith}, {McConnachie}, {Gwyn}, {Hayes}, {Gatto}, {Chambers}, {Cuillandre}, {Hudson}, {Magnier}, {Martin}, \& {Navarro}}]{Smith2025}
{Smith}, S. E.~T., {McConnachie}, A.~W., {Gwyn}, S., {et~al.} 2025, \apj, 983, 59

\bibitem[{{Sollima} {et~al.}(2011){Sollima}, {Mart{\'\i}nez-Delgado}, {Valls-Gabaud}, \& {Pe{\~n}arrubia}}]{Sollima2011}
{Sollima}, A., {Mart{\'\i}nez-Delgado}, D., {Valls-Gabaud}, D., \& {Pe{\~n}arrubia}, J. 2011, \apj, 726, 47

\bibitem[{{Sollima} {et~al.}(2012){Sollima}, {Nipoti}, {Mastrobuono Battisti}, {Montuori}, \& {Capuzzo-Dolcetta}}]{Sollima2012}
{Sollima}, A., {Nipoti}, C., {Mastrobuono Battisti}, A., {Montuori}, M., \& {Capuzzo-Dolcetta}, R. 2012, \apj, 744, 196

\bibitem[{{Stetson}(1987)}]{stetson87}
{Stetson}, P.~B. 1987, \pasp, 99, 191

\bibitem[{{Tanvir} {et~al.}(2012){Tanvir}, {Mackey}, {Ferguson}, {Huxor}, {Read}, {Lewis}, {Irwin}, {Chapman}, {Ibata}, {Wilkinson}, {McConnachie}, {Martin}, {Davies}, \& {Bridges}}]{Tanvir2012}
{Tanvir}, N.~R., {Mackey}, A.~D., {Ferguson}, A.~M.~N., {et~al.} 2012, \mnras, 422, 162

\bibitem[{{Taylor} {et~al.}(2025){Taylor}, {Read}, {Orkney}, {Kim}, {Pontzen}, {Agertz}, {Rey}, {Andersson}, {Collins}, \& {Yates}}]{Taylor2025}
{Taylor}, E.~D., {Read}, J.~I., {Orkney}, M. D.~A., {et~al.} 2025, \nat, 645, 327

\bibitem[{{Veljanoski} {et~al.}(2014){Veljanoski}, {Mackey}, {Ferguson}, {Huxor}, {C{\^o}t{\'e}}, {Irwin}, {Tanvir}, {Pe{\~n}arrubia}, {Bernard}, {Fardal}, {Martin}, {McConnachie}, {Lewis}, {Chapman}, {Ibata}, \& {Babul}}]{Veljanoski2014}
{Veljanoski}, J., {Mackey}, A.~D., {Ferguson}, A.~M.~N., {et~al.} 2014, \mnras, 442, 2929

\bibitem[{{Weisz} {et~al.}(2026){Weisz}, {Savino}, \& {Skillman}}]{Weisz2026}
{Weisz}, D.~R., {Savino}, A., \& {Skillman}, E.~D. 2026, arXiv e-prints, arXiv:2610.36040

\bibitem[{{Willman} {et~al.}(2005{\natexlab{a}}){Willman}, {Blanton}, {West}, {Dalcanton}, {Hogg}, {Schneider}, {Wherry}, {Yanny}, \& {Brinkmann}}]{Willman2005a}
{Willman}, B., {Blanton}, M.~R., {West}, A.~A., {et~al.} 2005{\natexlab{a}}, \aj, 129, 2692

\bibitem[{{Willman} {et~al.}(2005{\natexlab{b}}){Willman}, {Dalcanton}, {Martinez-Delgado}, {West}, {Blanton}, {Hogg}, {Barentine}, {Brewington}, {Harvanek}, {Kleinman}, {Krzesinski}, {Long}, {Neilsen}, {Nitta}, \& {Snedden}}]{Willman2005b}
{Willman}, B., {Dalcanton}, J.~J., {Martinez-Delgado}, D., {et~al.} 2005{\natexlab{b}}, \apjl, 626, L85

\bibitem[{{Zucker} {et~al.}(2006){Zucker}, {Belokurov}, {Evans}, {Wilkinson}, {Irwin}, {Sivarani}, {Hodgkin}, {Bramich}, {Irwin}, {Gilmore}, {Willman}, {Vidrih}, {Fellhauer}, {Hewett}, {Beers}, {Bell}, {Grebel}, {Schneider}, {Newberg}, {Wyse}, {Rockosi}, {Yanny}, {Lupton}, {Smith}, {Barentine}, {Brewington}, {Brinkmann}, {Harvanek}, {Kleinman}, {Krzesinski}, {Long}, {Nitta}, \& {Snedden}}]{Zucker2006}
{Zucker}, D.~B., {Belokurov}, V., {Evans}, N.~W., {et~al.} 2006, \apjl, 643, L103

\end{thebibliography}

\begin{appendix}

\section{Supplementary material}

\begin{minipage}{\textwidth}
\includegraphics[width=0.9\textwidth]{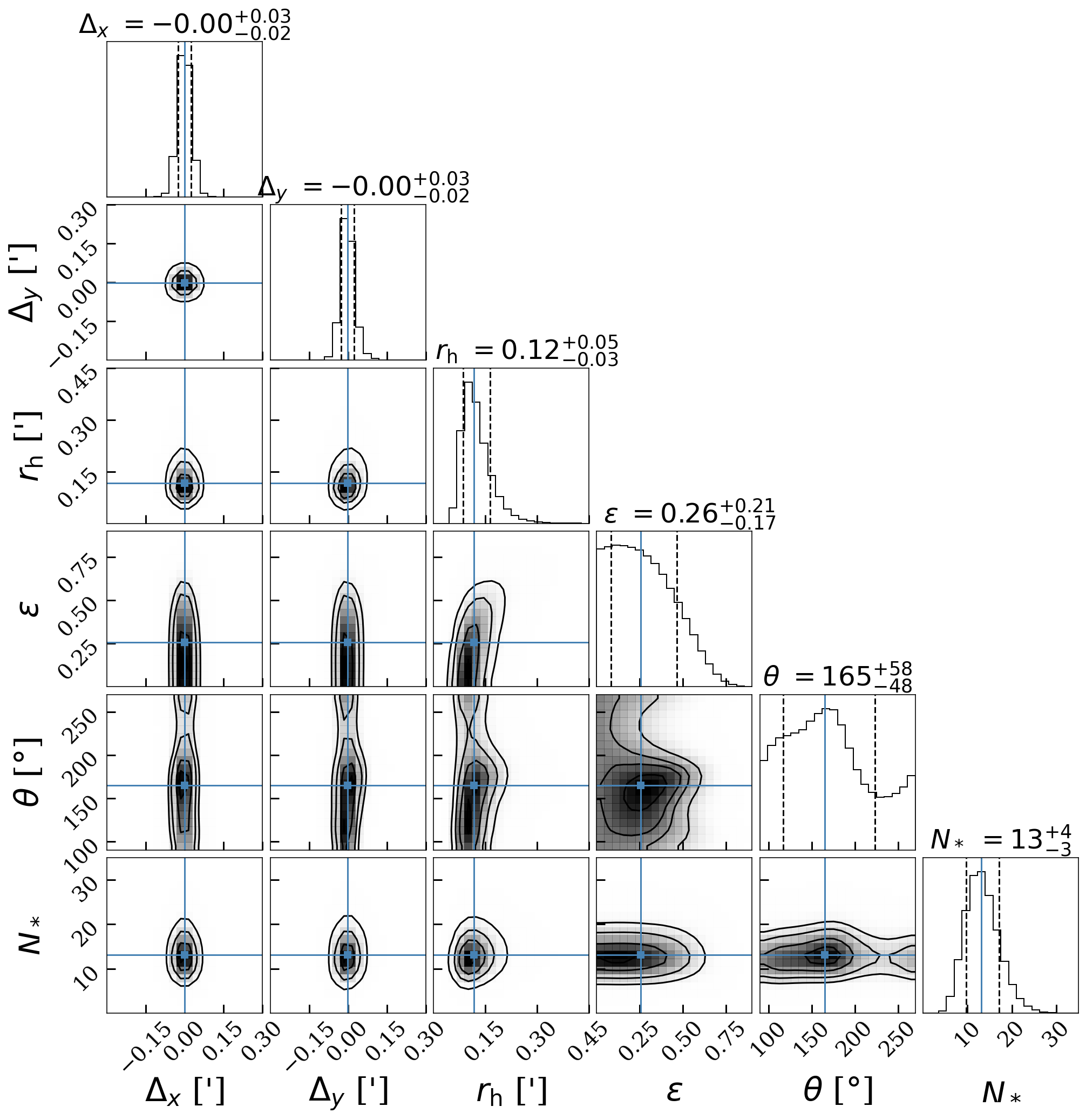}
\captionof{figure}{Results of the MCMC structural parameter analysis described in Sect. \ref{sec:res}. Bold lines mark the median parameter value and dashed lines the 16th and 84th quantile. $\Delta x$, $\Delta y$ denote the shift (in arcmin) in the assumed sky position of the overdensity's centre (Fig. \ref{fig:CMD}, right panel). $r_{\rm h}$ is the half-light radius (arcmin), $\epsilon$ is the ellipticity, $\theta$ is the position angle (degrees), and $N_{*}$  is the number of stars. }
\label{fig:mcmc}
\end{minipage}

\end{appendix}

\end{document}